\documentclass[10pt,journal,compsoc]{IEEEtran}
\ifCLASSOPTIONcompsoc
  \usepackage[nocompress]{cite}
\else
  \usepackage{cite}
\fi
\ifCLASSINFOpdf
   \usepackage[pdftex]{graphicx}
\else
\fi
\usepackage{amsmath}
\usepackage{mathtools}
\usepackage{amsfonts}
\DeclarePairedDelimiter{\ceil}{\lceil}{\rceil}
\usepackage{algorithmicx}
\usepackage{algpseudocode}
\usepackage[linesnumbered,titlenumbered,ruled,vlined,resetcount,algosection]{algorithm2e}
\makeatletter
\newcommand{\removelatexerror}{\let\@latex@error\@gobble}
\makeatother

\usepackage{tabularx}
\newcommand\setrow[1]{\gdef\rowmac{#1}#1\ignorespaces}
\newcommand\clearrow{\global\let\rowmac\relax}
\usepackage[color=pink]{todonotes}
\clearrow

\usepackage{listings}
\usepackage{omplst}
\usepackage{subcaption}
\usepackage[utf8x]{inputenc}
\lstdefinestyle{code}{
    basicstyle=\scriptsize,
    numbers=left,
    numberstyle=\tiny,
    numbersep=-3pt,
    stepnumber=1,
    frame=lines,
    stringstyle=\ttfamily,
    keywordstyle=\bfseries,
    tabsize=4,
    commentstyle=\color{gray}\tiny\sffamily\textit,
    columns=fixed
}
\lstdefinestyle{C} {
    style=code,
    language=C
}
\lstdefinestyle{OMP}{
    style=code,
    language=[OpenMP]C,
    emph=[1]{sequence\_of\_statements, a\_type, initial\_value},
    emphstyle=[1]\textit
}

\newcommand{\orc}{\includegraphics[height=\fontcharht\font`A]{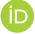}}
\usepackage[hidelinks]{hyperref}
\begin{document}
%
\title{Taskgraph: A Low Contention OpenMP Tasking Framework}
\author{Chenle Yu \href{https://orcid.org/0000-0002-1802-8680}{\orc}, Sara Royuela \href{https://orcid.org/0000-0002-7644-0868}{\orc}, Eduardo Quiñones \href{https://orcid.org/0000-0002-5465-964X}{\orc}}

\IEEEtitleabstractindextext{%
\begin{abstract}
OpenMP is the de-facto standard for shared memory systems in High-Performance
Computing (HPC). It includes a task-based model that offers a high-level of
abstraction to effectively exploit highly dynamic structured and unstructured 
parallelism in an easy and flexible way. Unfortunately, the run-time overheads 
introduced to manage tasks are (very) high in most common OpenMP frameworks 
(e.g., GCC, LLVM), which defeats the potential benefits of the tasking model, 
and makes it suitable for coarse-grained tasks only.
This paper presents \texttt{taskgraph}, a framework that uses a task dependency
graph (TDG) to represent a region of code implemented with OpenMP tasks in 
order to reduce the run-time overheads associated with the management of tasks, 
i.e., contention and parallel orchestration, including task creation and 
synchronization. The TDG avoids the overheads related to the resolution of task 
dependencies and greatly reduces those deriving from the accesses to shared 
resources. Moreover, the taskgraph framework introduces in OpenMP the 
\textit{record-and-replay} execution model that accelerates the 
taskgraph region from its second execution. Overall, the multiple optimizations 
presented in this paper allow exploiting fine-grained OpenMP tasks to cope 
with the trend in current applications pointing to leverage massive on-node 
parallelism, fine-grained and dynamic scheduling paradigms.
The framework is implemented on LLVM 15.0.
Results show that the taskgraph implementation outperforms the vanilla OpenMP 
system in terms of performance and scalability, for all structured and 
unstructured parallelism, and considering coarse and fine grained tasks. 
Furthermore, the proposed framework considerably reduces the performance gap 
between the task and the thread models of OpenMP.
\end{abstract}

}

\maketitle

\IEEEdisplaynontitleabstractindextext

%
\IEEEpeerreviewmaketitle


\textit{This work has been submitted to the IEEE for possible publication. Copyright may
be transferred without notice, after which this version may no longer be accessible.}

\section{Introduction}
\label{sec:intro}

OpenMP has been the standard in shared-memory parallel programming over the past 
decades. It allows users to parallelize programs by simply inserting 
annotations. Although initially the specification only targeted data-parallelism 
by means of the \textit{thread model}, from version 3.0 it has grown to 
support unstructured and highly dynamic parallelism through the \textit{tasking 
model}, and to support accelerators (from version 4.0) extending the tasking 
model for offloading work to the device. 

The tasking model offers a natural way to expose parallelism as it defines 
\textit{what} can be done in parallel, rather than defining \textit{how} to 
exploit parallelism (e.g., scheduling, mapping, etc.). Consequently, the OpenMP 
specification is evolving from a prescriptive thread model into a descriptive 
tasking model. Despite this trend, the most widely accepted OpenMP 
implementations of the tasking model, like GCC and LLVM, still fail at providing 
comparable performance to the thread model for structured applications. Previous 
works \cite{podobas2010comparison,podobas2016towards,rico2019benefits} have 
concluded that the overheads introduced by the runtime to manage tasks are too 
high, and so the potential of the OpenMP tasking model is reduced to the 
exploitation of rather coarse-grained tasks.

The reason is that the tasking model introduces several functionalities in the 
runtime (e.g., task creation, task scheduling and task synchronization) to 
orchestrate the parallel execution of tasks that may affect severely the 
performance. There are two main aspects influencing in the overhead in the 
OpenMP runtime 
\cite{furlinger2006analyzing,lagrone2011set,bull2012microbenchmark, 
gautier2018impact}: \textit{contention} and \textit{task granularity}.
The former occurs due to the simultaneous access to the system locks used to 
protect shared data structures, like the queues storing ready tasks, and it is
directly affected by the amount of parallelism exposed during the execution, 
the number of available threads and the work stealing mechanism when implemented.
The latter refers to the cost of managing 
the parallel execution with the respect to the total performance of the 
application; in this sense, the smaller the computation within a task, the 
bigger the impact of the overhead on the end-to-end execution time.

There are solutions that reduce the overheads of the runtime system aiming at 
the exploitation of fine-grained task-based parallelism with OpenMP (see 
Section~\ref{sec:related} for more details). These follow two main approaches: 
(1) using low-level threading solutions other than Pthreads, like user-level 
threads, typical from high-performance computing (HPC) domains 
\cite{Iwasaki2019}; and (2) tuning the application for a given architecture, 
typical from embedded computing (EC) domains. We propose a third approach based 
on the concept of Taskgraph, which is orthogonal to the other two, and 
can be built on top of them to further enhance the performance of each 
particular system. For this reason, the comparison of the aforementioned 
approaches with the proposed Taskgraph framework remains out of the scope of 
this work.

This paper addresses the run-time overhead introduced by the OpenMP tasking  
model by defining a new OpenMP task-based region, named \texttt{taskgraph}, 
that allows to transform a region of code where the computation is fully 
parallelized with OpenMP tasks into a task dependency graph (TDG). The 
TDG\footnote{In the following of this paper, we will use TDG and Taskgraph 
interchangeably, and we use \texttt{taskgraph} to refer to OpenMP related 
notations, e.g., a code region, a directive etc.} is a Directed Acyclic Graph 
(DAG) where nodes are task instances, and edges define dependencies among them.  
The TDG, which can either be generated statically, at compile-time, or 
recorded, at run-time when the region is executed for the first time, holds all 
the information required for the execution of the corresponding OpenMP region. 
The framework can hence replace the execution of the region with the execution 
of the TDG, reducing the run-time overheads due to contention and parallel 
orchestration, including task creation and synchronization, and so allowing the 
exploitation of finer-grained parallelism. 

The main contributions of this work are the following:
\begin{enumerate}
\item A novel solution, the Taskgraph framework, that involves analysis and 
optimization techniques based on the task dependency graph to alleviate the 
overheads of the OpenMP tasking model. The solution applies to both structured 
(e.g., \texttt{omp taskloop}) and unstructured (e.g., \texttt{omp task}) 
parallelism, and combines offline and online support to generate the TDG: If 
possible, the TDG is built at compile-time (a feature inspired on a previous 
work \cite{vargas2016lightweight}), to reduce the overhead from the first 
execution of the aimed OpenMP region; otherwise, a \textit{record-and-replay} 
method is used, to increase the efficiency from the second execution of the 
targeted region.
\item A complete implementation of the Taskgraph framework on Clang + LLVM 15.0 
compilation infrastructure \footnote{The implementation is uploaded at
https://anonymous.4open.science/r/llvm-tpds-E8F7} \cite{llvm15}. The compiler is
in charge of generating the code that either builds a static TDG, or generates
the calls to the runtime to record the TDG. The runtime counterparts implement 
the mechanisms needed to correctly record and execute the TDG. To further prove 
the framework's efficacy in other OpenMP implementations, we carried our work 
onto GCC 7.3.0 with Mercurium~\cite{bsc2020mercurium} as the front-end.
\item Complete evaluation of the proposed framework, in terms of runtime 
overhead, performance and scalability, considering different task granularities, 
number of threads and number of times the targeted regions are executed.
\end{enumerate}


\section{Motivation}
\label{sec:motiv}

Contention and parallel orchestration may have a great impact on the 
performance of OpenMP programs. To illustrate this effect, 
Listing~\ref{lst:omp_code} presents a synthetic benchmark that generates a 
bunch of series of independent tasks (embarrassingly parallel), where each task 
in a series depends on one unique task in the previous series. The 
corresponding TDG is depicted in Figure~\ref{lst:omp_code}.

\begin{figure}[h!]
  \centering
\begin{minipage}{0.95\columnwidth}
\begin{lstlisting}[style=OMP,caption=OpenMP tasking example.,label=lst:omp_code]
  #pragma omp parallel shared(deps)
  #pragma omp single
  {
    (*@\textit{START\_TIMER}@*)
    for (int i = 0; i < NUM_TASKS; i++) {
       #pragma omp task depend(out: deps[i % NUM_CORES])
       fn();
    }
    (*@\textit{END\_TIMER}@*)
  }
\end{lstlisting}
\end{minipage}

\begin{minipage}{0.97\columnwidth}
  \vspace{0.3cm}
  \centering
  \includegraphics[scale=0.6]{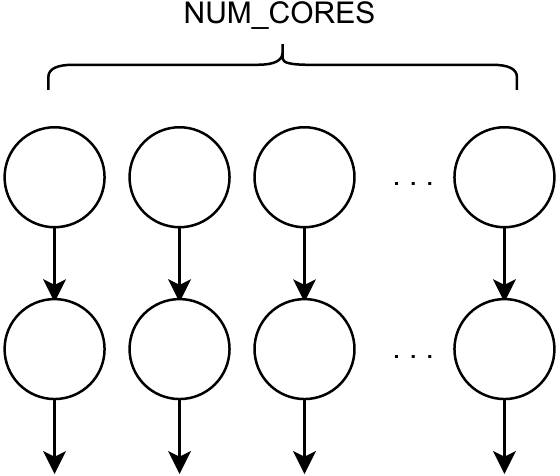}
  \caption{TDG representation of code in Listing~\ref{lst:omp_code}.}
\end{minipage}
\end{figure}

\begin{figure}[h!]
  \centering
  \begin{minipage}[t]{\columnwidth}
    
\includegraphics[height=160pt,width=.95\columnwidth]{
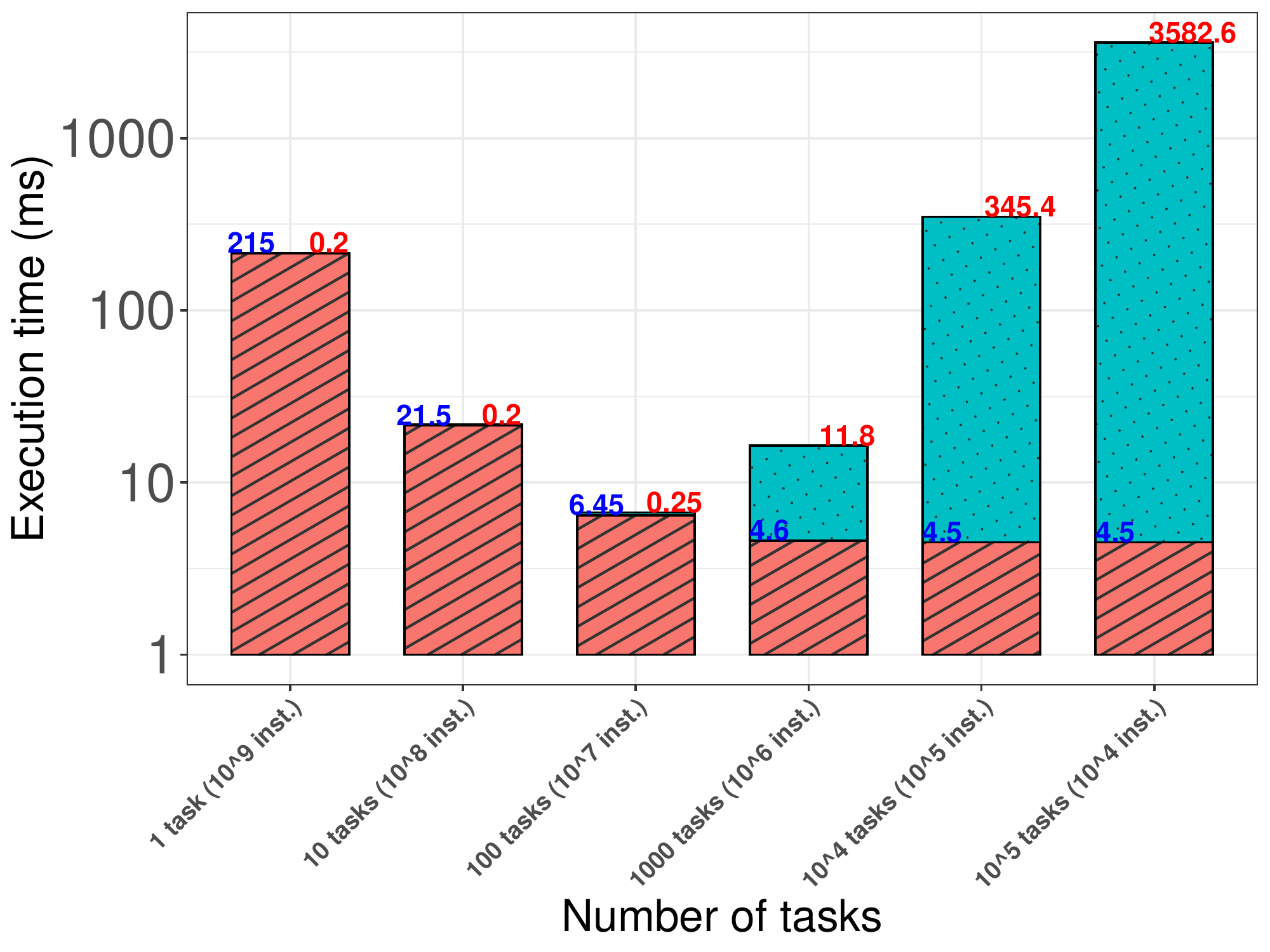}
    \subcaption{GCC GOMP runtime library.}
    \label{fig:gcc_over}
  \end{minipage}
  \begin{minipage}[t]{\columnwidth}
\includegraphics[height=160pt,width=.95\columnwidth]{
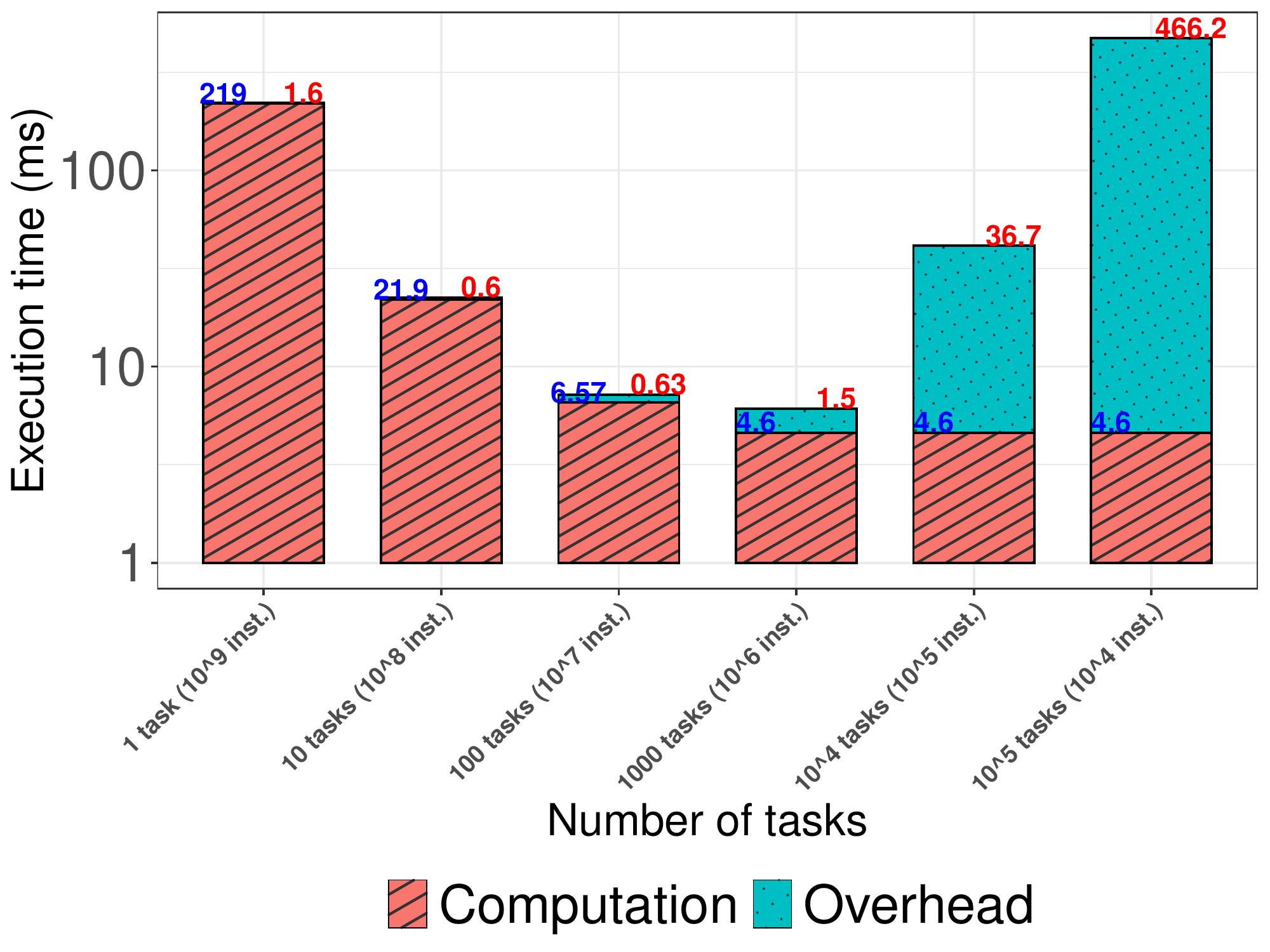}
    \subcaption{LLVM OpenMP runtime library.}
    \label{fig:llvm_over}
  \end{minipage}
  \caption{Task orchestration overhead in original GCC and LLVM tool-chains, 
           executing the program in listing~\ref{lst:omp_code} on a node of 
           Marenostrum 4~\cite{bsc2017mn4}.}
  \label{fig:runtime_over}
\end{figure}

The time measurements (as shown in Figure~\ref{fig:runtime_over}) only start 
after the \texttt{single} construct to eliminate the impact of the 
\texttt{parallel} and \texttt{single} directives in the evaluation of task 
overhead. Additionally, for this experiment the total workload remains 
unchanged, and it is set to $10^9$ assembly instructions. These instructions are 
hence distributed evenly among tasks, meaning that task granularity decreases 
while the number of tasks increases.

The evaluation considers two metrics: $Computation$ and runtime $Overhead$. The 
former represents the optimal execution time if the OpenMP runtime was free of 
overhead. The latter is the difference between measured time and $Computation$. 
As tasks within each series are independent, we can calculate both metrics as 
the formulas below:

\begin{equation*}
  let\ Time_{fn} = \frac{Serial\_Time}{c(Ta)}
\end{equation*}
\begin{align}
  Computation &= Time_{fn} * \ceil{\frac{c(Ta)}{c(Th)}}\\
  Overhead    &= Measured\_time - Computation
\label{eq:overhead}
\end{align}

where $Serial\_Time$ the time needed to execute the program sequentially. 
$c(Ta)$ and $c(Th)$ are the numbers of tasks and threads, respectively.
$Time_{fn}$ is the time needed to execute $fn$ once.
$Measured\_time$ represents the time we measure using \texttt{START\_TIMER} and
\texttt{END\_TIMER}.

As the number of threads is also unchanged, while varying the number of tasks we 
can determine the optimal $Computation$ as follows:

Case 1, $ c(Ta) \leq c(Th) $:

\begin{equation*}
  \ceil{\frac{c(Ta)}{c(Th)}} = 1
\end{equation*}
\begin{align}
  Computation &= Time_{fn}
\end{align}

Case 2, $c(Ta) > c(Th)$:

\begin{equation*}
\begin{split}
  c(Ta) = d * c(Th) + r, \ d\subset \mathbb{N}^{+},\  0 \leq r < c(Th) \\
  \ceil{\frac{c(Ta)}{c(Th)}} = \ceil{d + \frac{r}{c(Th)}} = d + 1 \\
  Computation = Time_{fn} * (d + 1)
\end{split}
\end{equation*}

\begin{align}
  \centering
  \lim_{c(Ta) \to +\infty} Computation &= \frac{Serial\_Time}{c(Th)}
\end{align}

As a result, when the number of threads is fixed, $Computation$ reduces if the 
number of tasks increases. Furthermore, $Computation$ tends to be constant when 
$c(Ta)$ is largely greater than $c(Th)$.

The synthetic program has been compiled with with GCC 7.3.0 and LLVM 15.0, with
the -O3 optimization flag. The tests have been carried out on a node of the 
Marenostrum 4 Supercomputer~\cite{bsc2017mn4}, using all 48 physical cores 
available in a dual-socket Intel Xeon Platinum. The results of the experiment 
are reported in Figure~\ref{fig:runtime_over}.

The numbers indicate that the overhead increases when the amount of tasks 
becomes larger. In fact, large amount of fine-grained tasks make threads to 
access more frequently to the task queue(s), resulting in a higher contention 
and synchronization cost. This trend is more pronounced in the GCC OpenMP 
library (GOMP) because of two reasons. First, in GOMP, all threads in a team
share an unique task queue. LLVM, instead, implements a distributed system
that creates a task queue for each thread, incurring in less lock contention.
Second, LLVM uses fine-grained locking to access team-shared resources,
such as the task dependency tracking hash table, where each hash
entry has its own lock. Oppositely, GCC wraps the entire hash table within a
massive locking region.

There is a remarkable difference between LLVM and GOMP when executing more than 
$10^3$ tasks, e.g., for $10^5$ tasks, the former takes 3582.6ms while the latter 
requires 466.2ms. Although the LLVM implementation shows consistently less 
overhead than GOMP, making it much more efficient to handle fine-grained tasks, 
there is evidence that both OpenMP runtimes do suffer from the overhead 
increment when managing fine-grained tasks. These observations together with 
works mentioned in Section~\ref{sec:intro}, show the need for a low-overhead 
solution for the tasking model.

\section{Related Work}
\label{sec:related}

The overhead introduced by the OpenMP runtime to handle parallelism is a widely 
studied topic because of two main reasons: (1) the upcoming of the exascale era 
is boosting the increase of the number of cores per processor in order to support 
massive on-node parallelism, which also augments the run-time overheads 
\cite{iwainsky2015many}, and (2) the interest generated by the OpenMP framework 
in other domains, like embedded computing, where the resources are much more 
limited compared to HPC \cite{agathos2013deploying}.

S. Olivier et. al.~\cite{olivier2020} studied the efficiency of the OpenMP 
tasking model of various implementations on different architectures. The results 
show that many can achieve an efficiency of 90\% with coarse-grained tasks, but 
dealing with fine-grained tasks becomes challenging. Implementations
optimize their libraries individually, yielding an efficiency outcome
ranging from 40\% to 80\%.

The most popular OpenMP runtimes rely on kernel-level threads, i.e., POSIX, 
\cite{buttlar1996pthreads}, to implement parallelism. This technique has the 
advantage of giving the scheduler full knowledge about the threads used in the 
application, but may result in important overheads when it comes to thread 
creation, switching and synchronization. As an alternative, other runtimes try 
to exploit fine-grained and highly dynamic 
parallelism by using lightweight thread (LWT) solutions based on the 
implementation of user-level threads (ULTs), a lighter representation of a 
thread fully controlled by the runtime system. However, for this reason, the OS 
sees ULTs as single-threaded processes, and hence the scheduler may incur poor 
decisions. An application of such a technique is GLTO \cite{castello2017glto}, 
an OpenMP runtime built on top of the Generic Lightweight Threads (GLT) API 
\cite{castello2017glt}, that allows running three different LWT libraries: 
MassiveThreads \cite{nakashima2014massivethreads}, 
Qthreads\cite{wheeler2008qthreads} and Argobots\cite{seo2016argobots}. This 
approach has shown better performance compared to current solutions like GNU 
and Intel for fine-grained tasks and nested parallelism. Nevertheless, regular 
solutions still perform better for work-sharing constructs. Another example is 
the Nanos++ \cite{bsc2020nanox} runtime system, which uses user-level threads 
\cite{teruel2007implementation}.

In the scope of Embedded Computing (EC), implementations address fine-grained 
tasks for specific architectures. For example, a runtime targeting 
tightly-coupled clusters \cite{burgio2013enabling} that focuses on tuning the 
the work queue, the implementation of Task Scheduling Points (TSPs) 
and the way task descriptors are interconnected. Another example is the 
extension \cite{tagliavini2018unleashing} of the previous runtime targeting the 
Kalray MPPA \cite{de2015kalray}, which modifies aspects such as the policy to 
insert suspended tasks in the ready queue and implements support for untied 
tasks.
Although the latter work~\cite{tagliavini2018unleashing} shows better
scalability than the first one~\cite{burgio2013enabling} when reducing
task granularity, it is not portable if compared to several other HPC
runtimes, like GOMP, Intel TBB \cite{kukanov2007foundations}, and also EC 
runtimes, like the MPPA SDK \cite{kalraySDK} based on GNU GCC.

Finally, another approach consists on compiler analysis techniques to statically
determine the TDG \cite{vargas2016lightweight}. This work eliminates the 
overheads introduced at run-time for managing task dependencies, and further 
extensions \cite{munera2020towards} limit the amount of on-the-fly tasks and
eliminate the use of dynamic memory. Both proposals have been developed on top 
of the Mercurium compiler and the GNU libgomp runtime library.

Although LWT might show better performance in different scenarios, they require 
a considerable effort to port tools and still fail at providing the expected 
performance for data-parallel applications. Additionally, ad-hoc solutions might 
also show better performance in specific machines, but they lack portability and 
maintainability. Based on that, and the fact the last approach is orthogonal to 
those using LWT, as they can still be used as low-level threading mechanisms, 
this work focuses on the concept of TDG.

This paper enhances previous works 
\cite{vargas2016lightweight,munera2020towards} by: (1) transparently capturing 
the data consumed by each OpenMP task and replacing the user code with the TDG 
execution, (2) overcoming the limitations of computing the TDG statically with
a \textit{record then replay} mechanism, and (3) providing an enhanced 
implementation that delivers better performance and boosts extensibility, as it 
is based on LLVM. These enhancements enable us to alleviate the overhead of 
OpenMP tasking model in many cases, especially with fine-grained tasks and 
many-threads environments.


\section{The Taskgraph Framework}
\label{sec:proposal}

This section introduces the Taskgraph framework, a set of mechanisms combining 
programming model, compiler and runtime support, that aim at replacing the 
execution of a region of user code that is \textit{fully taskified} (i.e., a 
region that contains only tasks and statements which computation is 
deterministic across different executions of the region or do not have side 
effects on the tasks) with the execution of the corresponding TDG. The optimized 
execution enabled by the TDG eliminates the overheads related to the task 
instantiation and dependency resolution, and reduces the contention on the data 
structures of the runtime.

\begin{figure}[h!]
  \centering
  \includegraphics[scale=0.4]{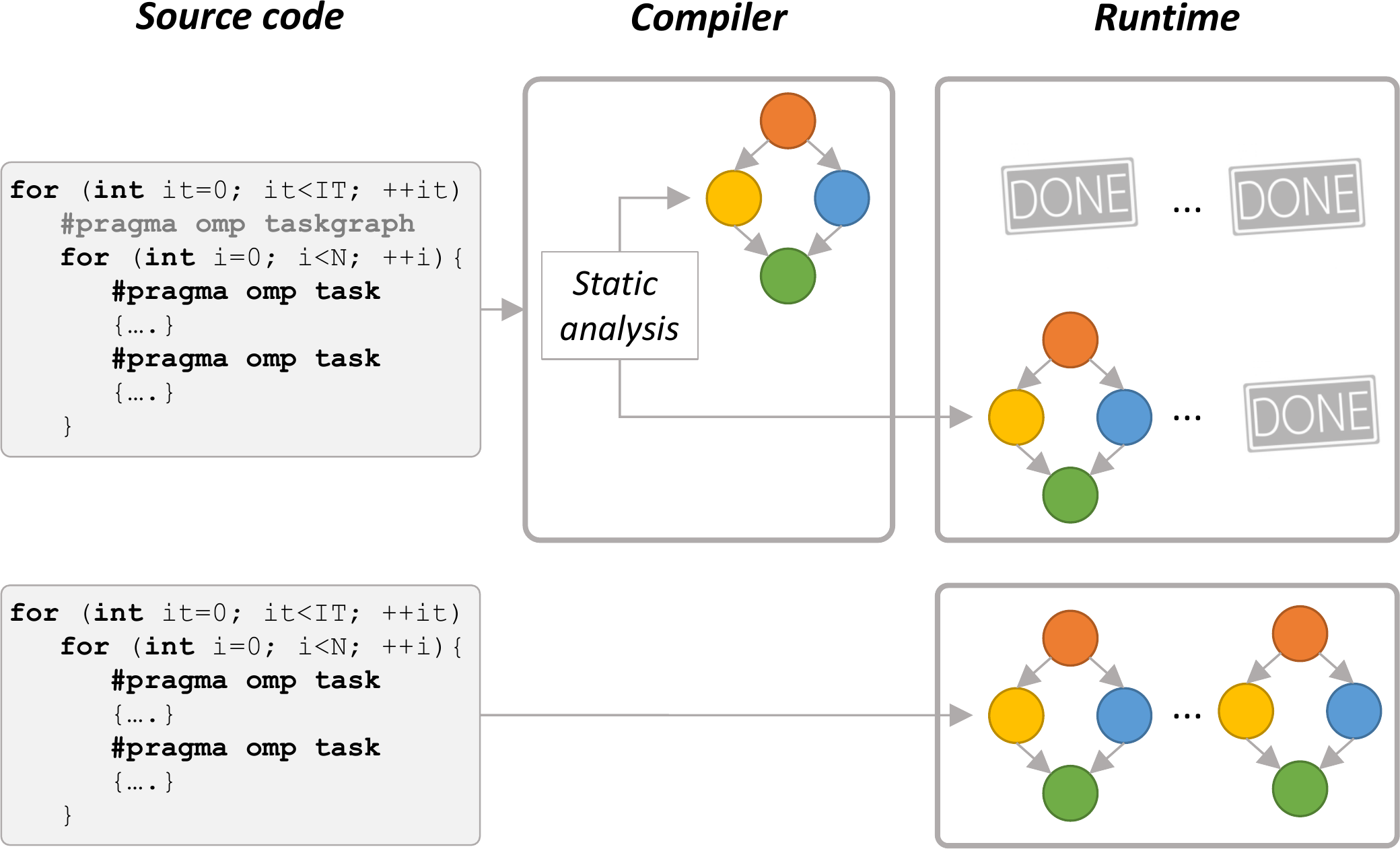}
  \caption{OpenMP frameworks overview: taskgraph (above) and vanilla (below).}
  \label{fig:taskgraph_framework}
\end{figure}

Figure~\ref{fig:taskgraph_framework} illustrates the work of the framework 
(above) compared to a regular OpenMP tool-chain (bellow), including programming 
model extensions, compilation tool, and run-time system. First (left), the 
\texttt{taskgraph} directive defines a region of code that can be fully captured 
by a TDG. Second (middle), the compilation tool performs static analysis on that 
region to decide whether the TDG can be generated at compile-time (this requires 
all variables involved in the generation of the tasks and their dependencies to 
be computable at compile-time), or it has to be recorded at run-time the first 
time the \texttt{taskgraph} region is called. 
Third (right), the run-time system is in charge of either (i) reading the TDG 
built by the compiler statically or; (ii) recording the TDG, by traversing the 
\texttt{taskgraph} region and saving the task structures and their dependencies. 
Once the TDG is loaded in the run-time system, independently of being built at 
compile-time or recorded at run-time, it can be re-executed any number of times.

\subsection{Programming model: Requirements}
\label{sec:requirements}
The region of code wrapped within a \texttt{taskgraph} directive has certain 
requirements, detailed as follows:
\begin{enumerate}
 \item All code in the \texttt{taskgraph} region must either (a) be enclosed 
within a task, or (b) belong to the control-flow statements in charge of 
instantiating tasks or do not cause side effects on the tasks.
 \item All code in the \texttt{taskgraph} region not included in a task must be 
constant across iterations, i.e., the shape of the TDG cannot change.
 \item The \texttt{taskgraph} directive cannot be declared recursively. In such 
cases, the inner TDG is fully encompassed by the outer TDG, and so defining the 
former becomes pointless.
\end{enumerate}
Based on the OpenMP philosophy, a code that uses the \texttt{taskgraph} 
directive and does not fulfill these conditions is non-conforming and it will 
result in unspecified behavior. To avoid this situation, the compilation tools
can perform the analysis needed to determine the conditions are met.

\subsection{Compiler: Analysis and transformations}
\label{sec:compiler_trans}

The use of static data-flow analysis enables an implementation of the 
Taskgraph framework that is transparent to the user, i.e., the framework decides 
if and which version of the Taskgraph can be used for each \texttt{taskgraph} 
directive. This section presents (1) the data-flow analysis required in the 
compiler, (2) the mechanisms that transparently generate the adequate Taskgraph 
version, and (3) the augmentations to support the \textit{record-and-replay} 
execution model. 

The static analyses and transformations have been implemented on top of two 
compilation frameworks: (1) the LLVM \cite{llvm15} compilation infrastructure, 
and (2) the Mercurium \cite{bsc2020mercurium} source-to-source compiler, 
targeting the GCC OpenMP runtime library \cite{gnu2020gomp}. 
Algorithm~\ref{algo:compiler_transformation} summarizes the technique 
to generate a Taskgraph, further explained in the remainder of the section.

\begin{figure}[h!]
\removelatexerror
\centering
\begin{algorithm}[H]\small
\KwData{TG, A taskgraph region.}
\KwResult{Transformation to be applied.}
 regular\_lowering = \KwSty{true}\;
 \eIf{control flow and dependencies known \\\hspace{8pt} $\&\&$ taskgraph 
restrictions met}{
     statically generate TDG \FuncSty{tdg\_src}\;
     \eIf{instance data known}{
       insert data in the TDG\;
        replace = \FuncSty{execute\_TDG(tdg\_src)}\;
     }   
     {   
       \eIf{recurrent\_taskgraph $\&\&$ first\_time \\\hspace{8pt} $||$ 
!recurrent\_taskgraph}
       {
          replace.append(\FuncSty{fill\_data()})\;
          replace.append(\FuncSty{execute\_TDG(tdg\_src)})\;
       }
       { replace =\FuncSty{execute\_TDG(tdg\_src)}}
     }
     \KwSty{TG} $\rightarrow$ replace\;
 }
 {
     \eIf{recurrent\_taskgraph}{
         \eIf{first\_time}{
             \KwSty{TG} $\rightarrow$ \FuncSty{record\_TDG(tdg\_fn)}\;
         }
         {
             \KwSty{TG} $\rightarrow$ \FuncSty{execute\_TDG(tdg\_src)}\;
         }
     }   
     {   
         \lForEach{task $t \in TG$}
	 {\\\hspace{8pt}$t$ $\rightarrow$ \FuncSty{instantiate\_task(...)}}
     }   
 }
 \caption{Compiler transformation mechanism for an OpenMP taskgraph region.}
 \label{algo:compiler_transformation}
\end{algorithm}
\end{figure}

\subsubsection{Data-flow analysis}

The mechanisms required at compile-time to generate a TDG from a \texttt{taskgraph}
region are based on control-flow and data-flow analyses. Among the former, the 
control flow graph (CFG) is the technique used to understand the flow of the 
application. Among the latter, loop unrolling and constant propagation are key 
to reveal the memory state each time a task instance is generated. 

The implementation slightly change from Mercurium to LLVM, mainly because 
Mercurium is a research compiler with a limited analysis infrastructure. 
Oppositely, LLVM provides a large set of analysis that allows for covering a 
wider range of the C/C++ standards. Nonetheless, both follow a similar approach 
and we use the LLVM pipeline as an illustration. First, the loop unrolling pass 
unrolls all loops involved in the instantiation of tasks. Then, the dominator 
tree pass is used to order the instructions, and determine the order of 
creation of tasks. After that, each task is assigned an unique identifier based 
on the order of creation. Once all tasks are discovered, the new TDG analysis 
calculates the dependencies among the tasks, obtaining a predecessor and 
successor list for each task instance, to be finally transformed into a task 
dependency graph. Furthermore, the analysis captures, if possible, the static 
data values associated with each task, this means, the values of the variables 
used inside the task at the time of creation.

\subsubsection{Generating the suitable single Taskgraph}
\label{sec:lower_TDG}

Based on the knowledge the compiler can gather, different lowering options are 
considered, reported in Figure~\ref{fig:omp_lowering}: the original code, Figure
 \ref{code:omp_src_par_reg}, shows an OpenMP application creating two tasks 
inside a loop; \ref{code:omp_src_orig} corresponds to the original lowering; 
\ref{code:omp_src_proposed_unknown} shows the proposed lowering when the TDG 
can be statically determined, but the data for each task instance cannot be 
computed; and \ref{code:omp_src_proposed_known} presents the proposed lowering 
when all data can be statically computed.

\begin{figure}[h!]
\centering
\begin{minipage}{.5\columnwidth}
\begin{subfigure}{1\textwidth}
\centering
\begin{lstlisting}[style=OMP,numbers=none]
 for (k=0; k<NB; k++) {
    #pragma omp task ...
    {...}
    #pragma omp task ...
    {...}
 }
\end{lstlisting}
\vspace{-.2cm}
\caption{Source parallel region.}
\label{code:omp_src_par_reg}
\end{subfigure}
\begin{subfigure}{1\textwidth}
\vspace{.2cm}
\begin{lstlisting}[style=OMP,numbers=none]
 fill_data(...);
 execute_TDG();
\end{lstlisting}
\caption{Lowering if unknown data.}
\label{code:omp_src_proposed_unknown}
\end{subfigure}
\end{minipage}
\hfill
\begin{minipage}{.48\columnwidth}
\begin{subfigure}{1\textwidth}
\vspace{.3cm}
\begin{lstlisting}[style=OMP,numbers=none]
 for (k=0; k<NB; k++) {
    instantiate_task(...);
    instantiate_task(...);
 }
\end{lstlisting}
\vspace{.11cm}
\caption{Original lowering.}
\vspace{.2cm}
\label{code:omp_src_orig}
\end{subfigure}
\begin{subfigure}{1\textwidth}
\vspace{.1cm}
\begin{lstlisting}[style=OMP,numbers=none]
 execute_TDG();
\end{lstlisting}
\vspace{.15cm}
\caption{Lowering if known data.}
\label{code:omp_src_proposed_known}
\end{subfigure}
\end{minipage}
\caption{High-level description of the lowering options available for 
\textit{taskgraph} regions, resulting from 
Algorithm~\ref{algo:compiler_transformation}.}
\label{fig:omp_lowering}
\end{figure}

As shown in Algorithm~\ref{algo:compiler_transformation}, the compiler can 
generate the TDG when all data used in the control-flow (i.e., loops and 
selection statements) that instantiates tasks and the data deciding the 
dependency clauses is known at compile-time, as long as the requirements 
described in section~\ref{sec:requirements} are met. Otherwise, the regular 
lowering, i.e., from \texttt{task} constructs to $instantiate\_task$ runtime 
calls, is used. Furthermore, when both the TDG and the data consumed by the 
tasks is known, this data is inserted in the TDG, and the execution of the user 
code is replaced by the execution of the TDG. If not, the data has to be 
captured at run-time, and once the TDG is complete, it can be executed. In this 
case, the execution of the user code can still be avoided by outlining the 
\texttt{taskgraph} region with an argument referencing all variables required 
for the execution of that region. Then, each task includes a series of pointers 
relative to the argument of the outlined function. At execution, tasks access
to the needed data by dereferencing their pointers.

\subsubsection{Generating the recurrent Taskgraph}
\label{sec:lower_taskgraph}
The transformations needed to implement the \textit{record-and-replay} 
execution model, shown in Algorithm~\ref{algo:compiler_transformation}, require 
the compiler to compute the \textit{call graph} of the application to decide if 
the \texttt{taskgraph} region is called recurrently. If it is recurrent, the TDG 
is recorded when the tasks are instantiated for the first time, before replacing 
subsequent calls by the execution of the TDG.

\subsection{Runtime: Orchestrating parallelism}
\label{sec:runtime_impl}
The runtime support for the Taskgraph is implemented on top of LLVM 15.0's 
KMP library \cite{llvm15} and GCC 7.3.0's OMP library. Again, we focus on
KMP for illustrative purposes. New structures and functions have been added 
in order to (1) hold the TDG, (2) complete the undetermined data in the static 
TDG if it is unknown at compile-time, (3) record a TDG, and (4) execute a 
previously created TDG, either statically or recorded.

\subsubsection{Using all threads' queue}
\begin{figure}[h!]
  \centering
    \begin{subfigure}{1\columnwidth}
      \centering
\includegraphics[width=0.95\columnwidth]{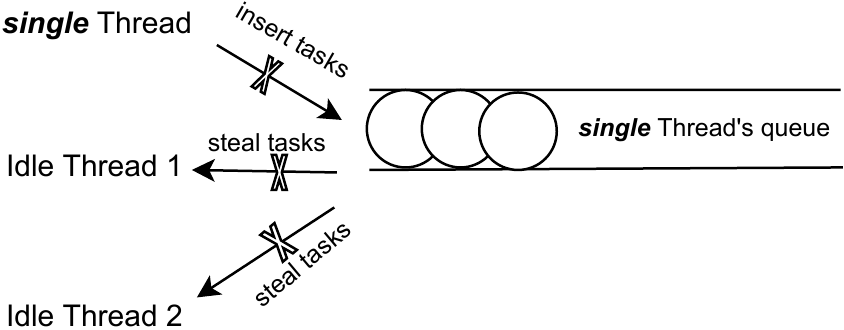}
      \caption{Current thread task queue interaction}
      \label{fig:taskqueue_before}
    \end{subfigure}
    \begin{subfigure}{1\columnwidth}
      \vspace{0.3cm}
      \centering
\includegraphics[width=0.95\columnwidth]{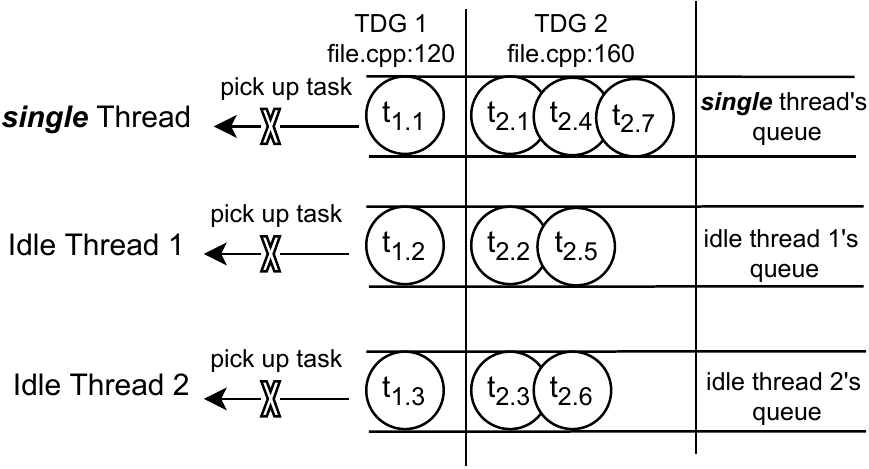}
      \caption{Thread task queue interaction once the TDG is built}
      \label{fig:taskqueue_after}
    \end{subfigure}
    \caption{LLVM Thread task queue representation. Arrows with "X" refer to 
exclusive
    \label{fig:taskqueues}
accesses}
\end{figure}

In most OpenMP task applications, only one thread executes the region of code 
spawning tasks (e.g., the thread entering \textit{single} construct). It then 
inserts these tasks into its own queue. Once its queue is not empty, this thread 
will notify idle threads to execute newly generated tasks. Once awake, idle 
threads start to steal tasks from \textit{single} thread's queue, which is 
protected by a lock, resulting in additional overhead due to contention. 
Figure~\ref{fig:taskqueue_before} depicts this scenario. To alleviate
this overhead, we consider all root tasks (i.e., tasks without input dependencies)
of a TDG, and evenly distribute them to different threads' queues. By doing so,
when executing a TDG, all threads will immediately pick up tasks from their 
own queue and start working, as represented in Figure~\ref{fig:taskqueue_after}.
This method is especially beneficial for applications with several tasks 
without input dependencies, like those exploiting the \texttt{taskloop} 
construct.

\subsubsection{TDG recording}
\label{sec:record_TDG}

When the TDG cannot be generated statically, the compiler emits a call to 
$record\_TDG$. This function executes the corresponding \texttt{taskgraph} 
region, while transparently records all tasks and their dependencies. The 
entries in the hash table tracking task dependencies are never freed. This way, 
dependencies with already finished tasks can be established, independently from 
the order of execution during the recording phase, and hence a correct TDG 
containing all necessary edges can be built. 

As soon as the TDG is built, root tasks can be evenly distributed among threads'
queues. The current implementation hands out tasks in a round-robin way. This 
decision is based on the aim to minimize the overhead. Although it may cause the 
total execution time of the assigned tasks to be very different among threads, 
the work-stealing mechanism can help in amortizing the load-imbalance. For 
instance, in TDG 2 from Figure~\ref{fig:taskqueue_after}, if the execution of 
task $t_{2.4}$ is unpredictably slow, idle threads 1 and 2 are allowed to steal 
task $t_{2.7}$ from the single thread once they completed their own work. Other 
solutions have been considered in order to enhance the load balance across 
threads, including the computation of the execution time of the tasks while 
recording. However, this incurred in larger overheads due to time measurement 
and scheduling logic.

\subsubsection{TDG execution}
\label{sec:exec_TDG}

Since there may be more than one TDG stored in the memory at a given time, we 
associate each TDG with their source location (i.e., file name and line number). 
To initiate the execution of a TDG, we invoke $execute\_TDG$ passing the source 
location of the TDG as an argument. Then, all root tasks in the TDG are 
inserted into a thread's queue. When a task finishes, it is in charge of 
inserting any successor if all its input dependencies are met.

Overall, the execution of the TDG does not require to allocate or free any data 
structure as all the information needed is accessible in the memory. Moreover, 
the TDG removes the need to exclusively access to the dependency-tracking hash 
table at each dependency resolution, because the work is already performed. 

By definition, a \texttt{taskgraph} region must contain all the information 
related to the execution of the tasks within that region. This means all the 
dependencies the tasks may have have to be represented, and so different 
instances of the same Taskgraph can only be executed in parallel if they do not 
contain dependencies among them. As this is hardly the case, the default 
behavior is to sequentialize the execution of different instances of the same 
TDG. Only when the \texttt{taskgraph} construct is augmented with a 
\texttt{nowait} clause, two instances can be executed in parallel. It remains as 
a future work to allow concurrent instances of the same TDG.

\section{Evaluation}
\label{sec:eval}

This section evaluates the proposed Taskgraph framework from different angles: 
(1) overhead with respect to the original runtime (preliminary analyzed in 
Figure~\ref{fig:runtime_over}); (2) speedup gain over the original 
implementations with respect to task granularity and thread number;
and (3) performance in repetitive execution model. For (1) and (2), we assume a 
TDG is already built, either statically or dynamically, and for (3), we take 
into account the cost of recording the TDG.

\subsection{Experimental setup}
\label{sec:experimentalSetup}
The experiments mainly focus on the implementation of the proposed framework on 
top of LLVM. However, results obtained with Taskgraph on top of GCC are also 
shown to strengthen the conclusions (see Sections~\ref{sec:over_reduc} and 
~\ref{sec:peak}).

One node of the Marenostrum 4 \cite{bsc2017mn4} has been used as processor. The 
node is equipped with 96 GB of RAM and dual-socket Intel Xeon Platinum 8160 
CPU. Each socket has 24 physical cores, with an L3 cache of 33 MB.

Some environment variables are set to reduce the impact of the access to the 
different sockets in the speed of the memory accesses, a.k.a. \textit{the NUMA 
effect}, as well as thread context switches, i.e., OMP\_PROC\_BIND=1, and 
GOMP\_CPU\_AFFINITY="0 - \$NUM\_THREADS-1", so each thread bounds to a physical 
core. 

A set of kernels and real-life applications from different domains have been 
used to evaluate the performance and applicability of the proposal. These are 
listed below: 
\begin{itemize}
\renewcommand\labelitemi{--}
\item A CPU bound synthetic application presented in Section~\ref{sec:motiv} to 
evaluate the contention overhead.
\item A \textit{Heat} diffusion simulator implemented with the Gauss-Seidel 
method, showing a Stencil computation. The matrix size of $2048 \times 2048$.
\item A \textit{Gravity force} simulator implemented with N-body algorithm, 
embarrassingly parallel. The size of particle set ranges from 512 to 4096.
\item \textit{AXPY}, a basic linear algebra operation that multiplies vector X 
by scalar A and adds vector Y to it. Vectors have $2^{27}$ elements.
\item A dot product, \textit{DOTP}, computation that multiplies two vectors to 
produce a scalar. Both vectors have $2^{27}$ entries.
\item \textit{Cholesky} matrix decompostion, a commonly used basic linear 
algebra subprogram. The matrix size is of $2048 \times 2048$.
\item \textit{HOG}, an object detector implemented with Histogram of Gradients. 
It uses images of $1920 \times 1080$ pixels, corresponding to Full HD 
resolution.
\item A subset of \textit{NAS Parallel Benchmarks (CG, BT, FT, SP, EP, LU)} 
implemented in C++~\cite{LOFF2021743}. The \textit{IS} kernel is not suitable 
for \textit{taskgraph + taskloop} because its correctness relies on the static 
scheduling supported by OpenMP threading model and thread private data, using 
\textit{threadprivate} directive. Similarly, the \textit{MG} kernel is not used 
because its task-spawning loop changes at every iteration, which does not meet 
the requirements (c.f. Section~\ref{sec:requirements}) to use 
\textit{taskgraph}.
\end{itemize}

All above benchmarks exploiting unstructured task-based parallelism are set to 
a fixed problem size, which is divided in different blocks. The block number is 
parameterized so that one can easily configure the task granularity (typically, 
each task performs the computation of a block). The NAS benchmark suite, 
instead, is meant to compile binaries with different problem sizes. To deal with 
it, we fix the number of tasks to be generated by using \textit{omp taskloop 
num\_tasks(N)}. This way, when the problem size varies, the same amount of tasks 
are created with different granularities. Finally, all benchmarks are 
compiled with the \textit{-O3} optimization flag, and execution times are 
calculated as the arithmetic mean of at least 20 executions.

\subsection{Overhead reduction}
\label{sec:over_reduc}

Motivated by the results shown in Figure~\ref{fig:runtime_over}, this section 
evaluates the overhead of tasking model after extending GCC and LLVM with 
support for the Taskgraph framework. We reuse the code in 
Listing~\ref{lst:omp_code} to compute the results reported in 
Table~\ref{tab:overhead_comp}.

\begin{table}[h!]
\centering
\caption{Tasking model overhead using the vanilla runtime systems and the 
corresponding Taskgraph extensions.}
  \label{tab:overhead_comp}
\resizebox{\columnwidth}{!}{%
  \begin{tabular}{|>{\rowmac}c |>{\rowmac} c |>{\rowmac} c |>{\rowmac} c |>{\rowmac} c |>{\rowmac} c |>{\rowmac} c<{\clearrow}|}
    \hline
      \begin{tabular}{c}
          $task\ number$ \\
          \hline
          $compiler$
      \end{tabular} 
             & $10^0$ & $10^1$ & $10^2$ & $10^3$ &$10^4$ &$10^5$ \\
    \hline
    GCC & 0.2 & 0.2 & 0.25 & 11.8 & 345.4 & 3582.6 \\
    \hline
    \setrow{\bfseries}GCC + taskgraph & 0.2 & 0.1 & 0.2 & 0.2 & 29.4 & 345.3 \\
    \hline
    LLVM & 1.6 & 0.6 & 0.6 & 1.5 & 36.7 & 466.2 \\
    \hline
    \setrow{\bfseries}LLVM + taskgraph & 0.2 & 0.1 & 0.2 & 0.3 & 9.9 & 132.2\\
    \hline
  \end{tabular}
}
  \vspace{1pt}
\end{table}

Both, GCC and LLVM, tool-chains show shorter overhead time after adopting the 
Taskgraph mechanism. In extreme cases, with $10^5$ small tasks of 10k 
instructions each, GCC + TDG incurs 10 times less overhead than the original GCC 
GOMP library (345.3ms compared to 3582.6ms, respectively). In the same trend, 
LLVM + TDG shows 3.5 times less overhead than the original LLVM OpenMP runtime 
library (132.2ms compared to 466.2ms, respectively).

This confirms the better suitability of the proposed framework to exploit 
fine-grained parallelism. One part of this overhead reduction comes from the 
Taskgraph minimizing the number of instructions executed to orchestrate tasks. 
This is so because all tasks are already allocated when the execution starts, 
and their dependencies are pre-computed. The other part of this overhead 
reduction is achieved by alleviating the contention on the shared resources. 
Each task creation and dependency resolution requires exclusive access to the 
shared resources (e.g. the task ready queue and the parent task's hash table). 
Consequently, when the number of available cores is high (48 in this case) and 
tasks show fine granularity, spinning to acquire the locks may result in 
enormous performance shrink. Table~\ref{tab:overhead_comp} reveals this 
scenario in the configurations with $10^4$ and $10^5$ tasks. 

\subsection{Performance gain with fine-grained task}
\label{sec:peak}

\begin{figure*}[h!]
  \centering
      \begin{minipage}[t]{\textwidth}
      \begin{subfigure}{.33\textwidth}%
        \includegraphics[width=\linewidth]{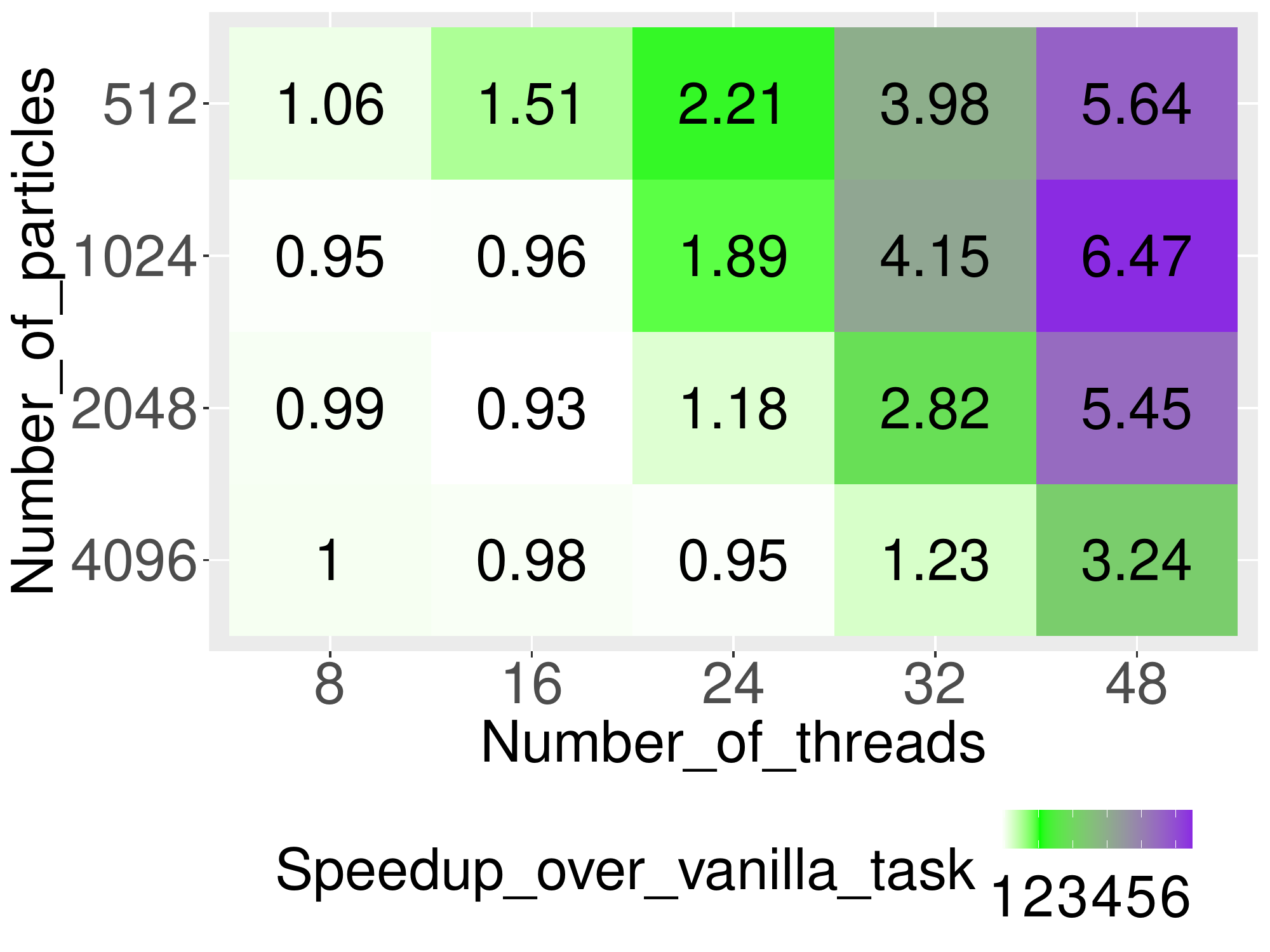}
        \vspace{-0.7cm}
        \caption{Nbody simulation}
        \label{fig:nbody_static_heatmap}
      \end{subfigure}
      \hfill
      \begin{subfigure}{.33\textwidth}%
        \includegraphics[width=\linewidth]{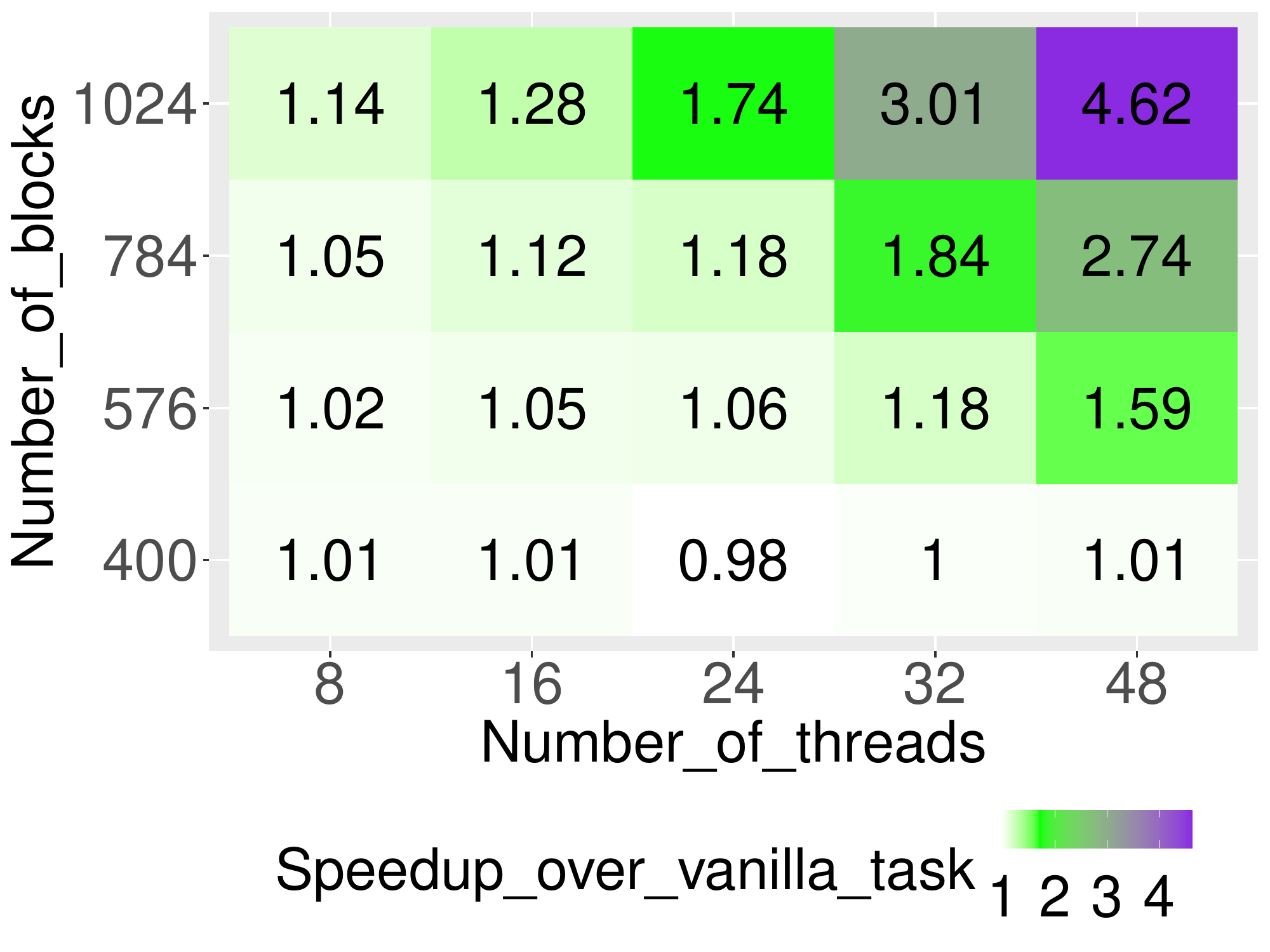}
        \vspace{-0.7cm}
        \caption{Cholesky decomposition}
        \label{fig:cholesky_static_heatmap}
      \end{subfigure}
      \hfill
      \begin{subfigure}{.33\textwidth}%
        \includegraphics[width=\linewidth]{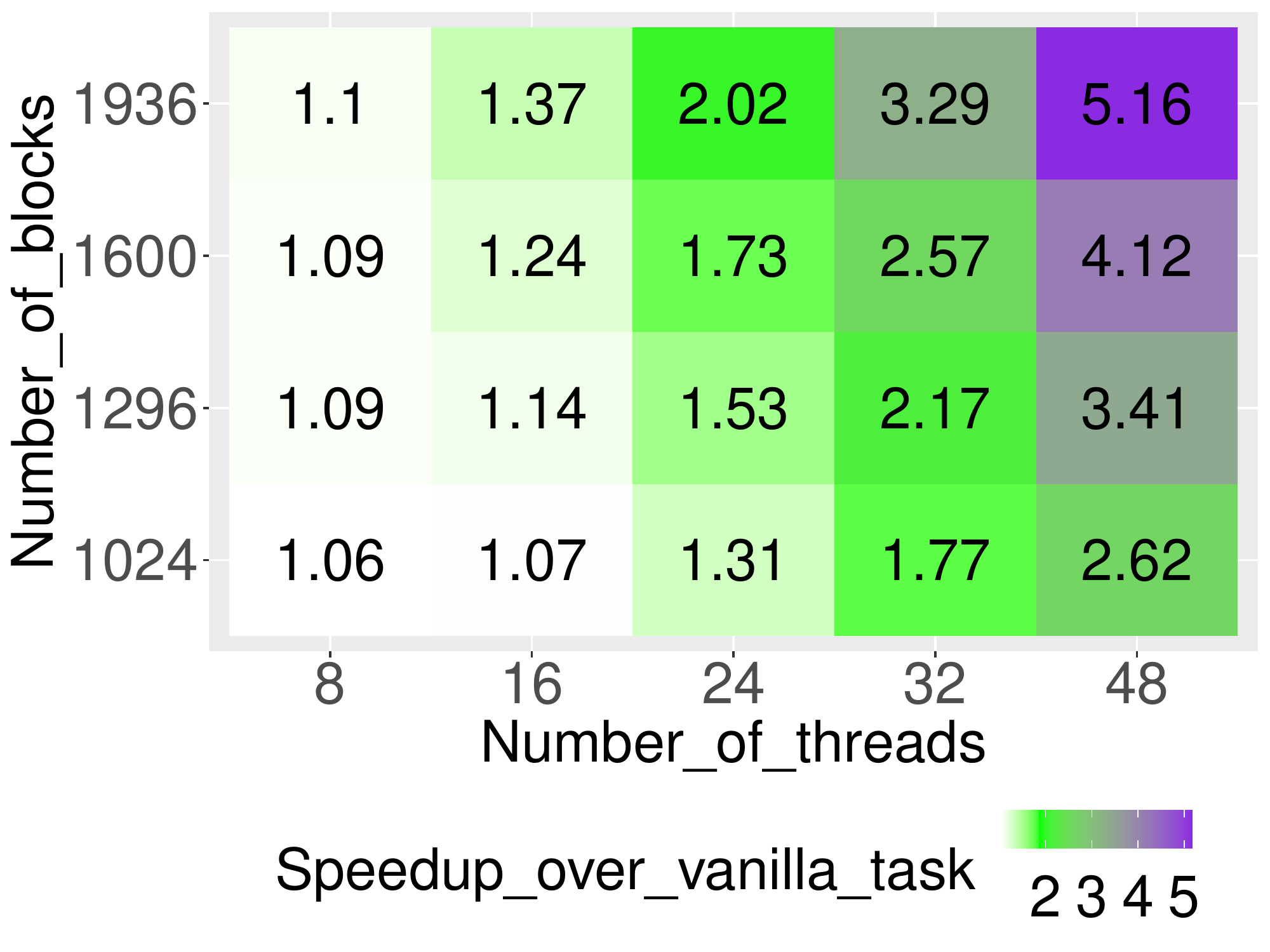}
        \vspace{-0.7cm}
        \caption{Heat propagation simulation}
        \label{fig:heatpropagation_static_heatmap}
      \end{subfigure}      
      \begin{subfigure}{.33\textwidth}%
        \vspace{0.3cm}
        \includegraphics[width=\linewidth]{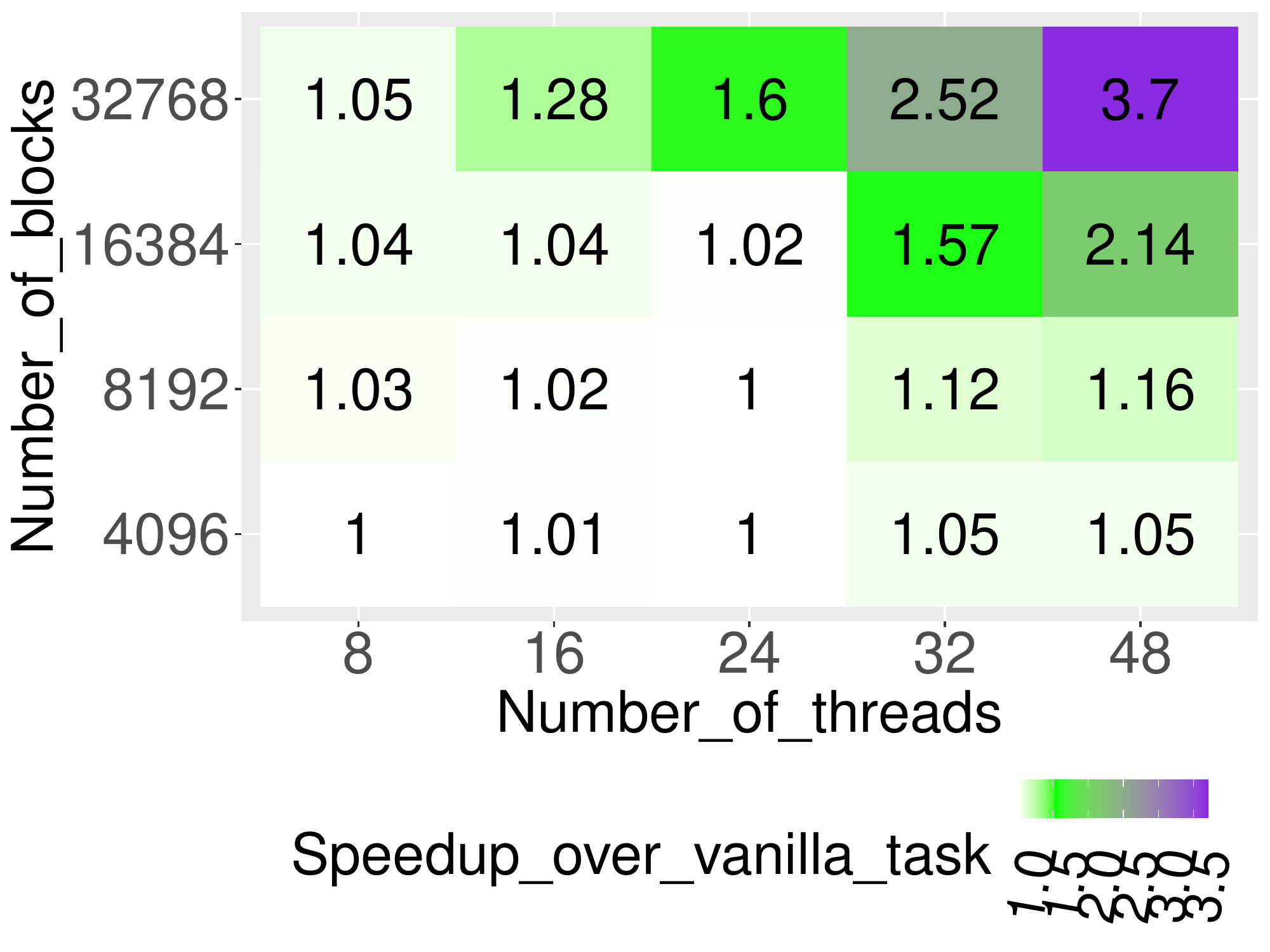}
        \vspace{-0.7cm}
        \caption{Dot product operation}
        \label{fig:dotp_static_heatmap}
      \end{subfigure}
      \hfill
      \begin{subfigure}{.33\textwidth}%
        \vspace{0.3cm}
        \includegraphics[width=\linewidth]{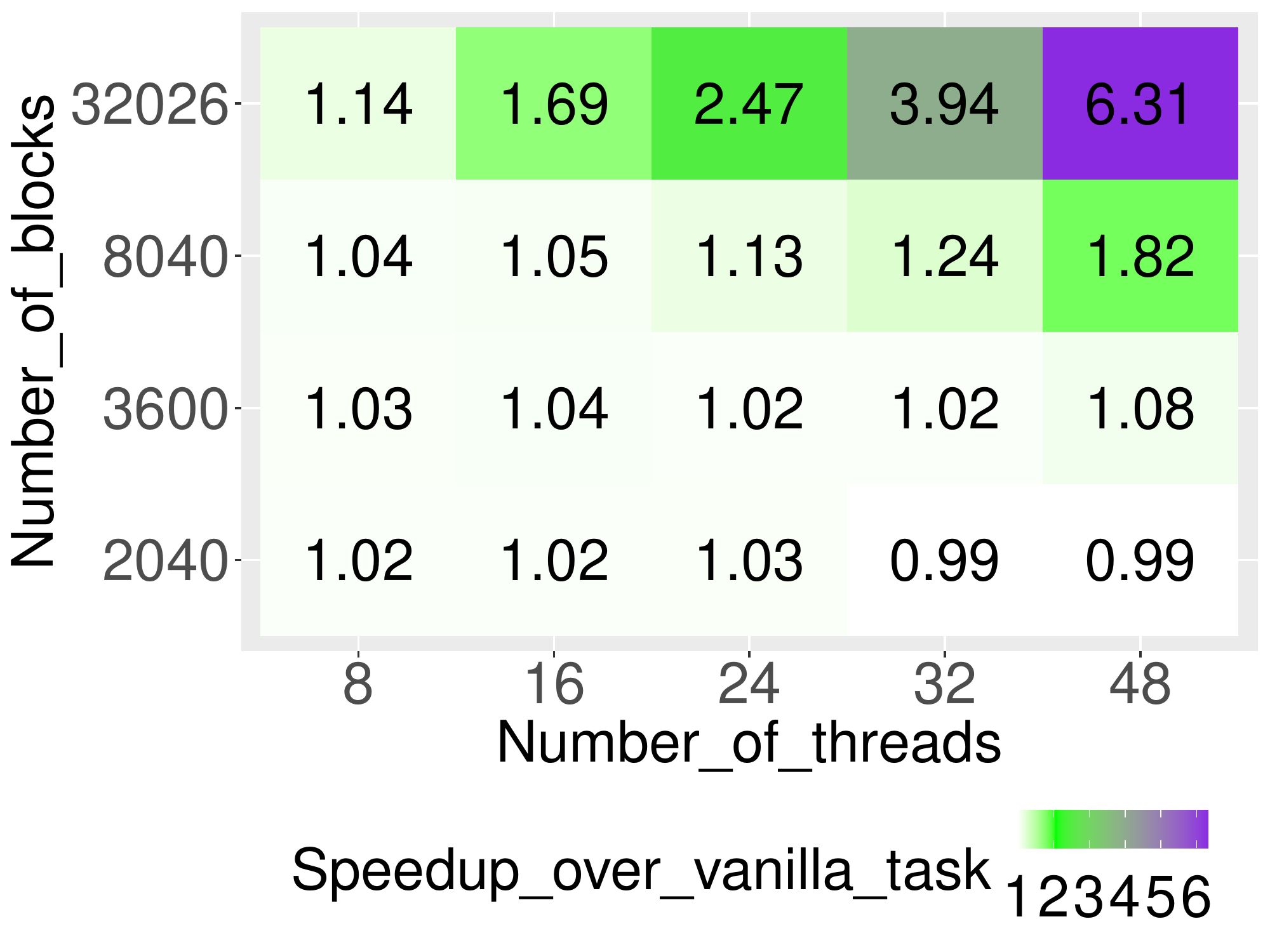}
        \vspace{-0.7cm}
        \caption{Hog object detector}
        \label{fig:hog_static_heatmap}
      \end{subfigure}
      \hfill
      \begin{subfigure}{.33\textwidth}%
        \vspace{0.3cm}
        \includegraphics[width=\linewidth]{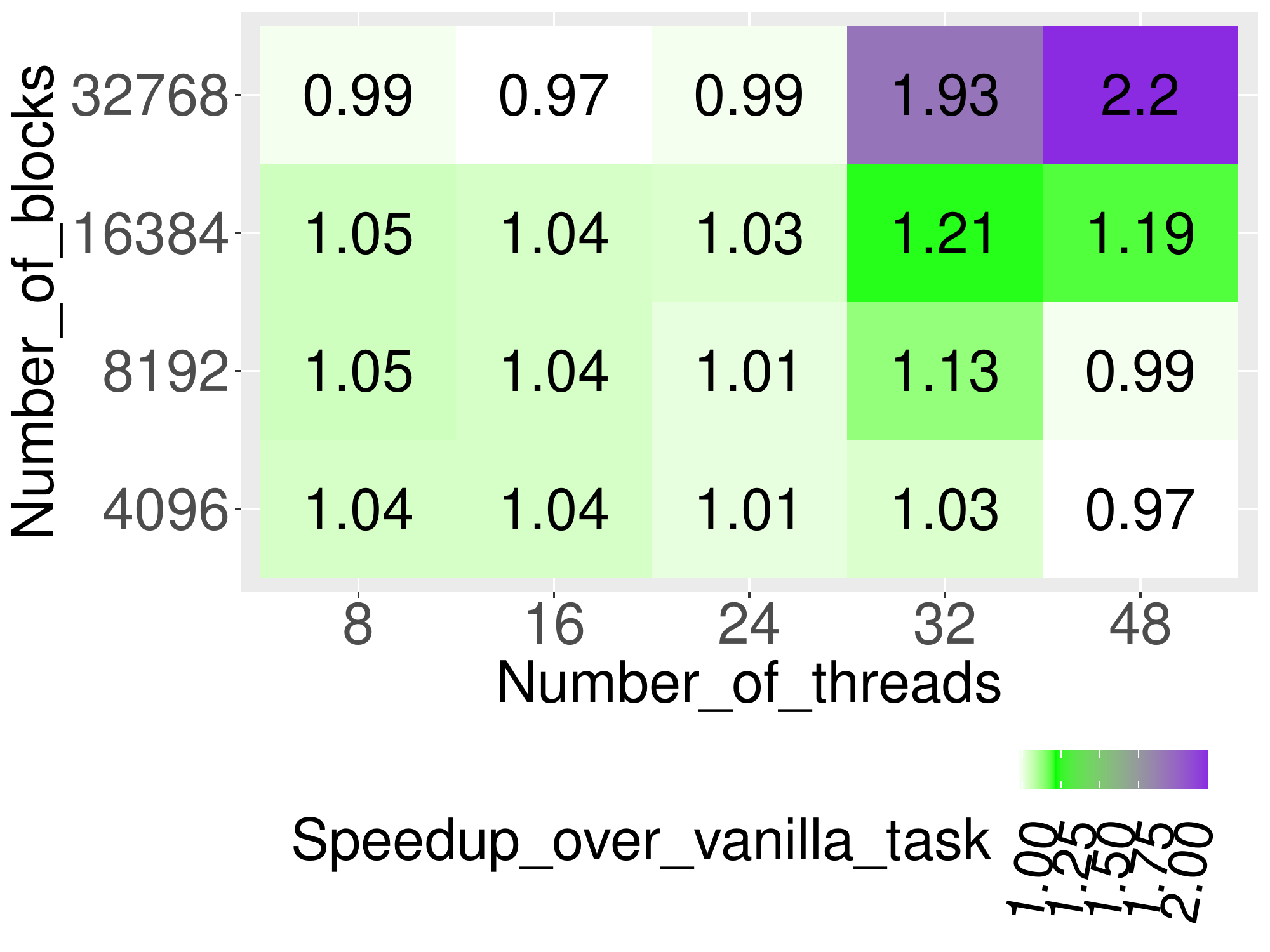}
        \vspace{-0.7cm}
        \caption{Axpy operation}
        \label{fig:axpy_static_heatmap}
      \end{subfigure}
      \vspace{-0.2cm}
      \caption{Normalized speedup of taskgraph compared to original \textit{task} construct.}
      \label{fig:unstruct_heatmaps}
      \end{minipage}
      \begin{minipage}[t]{\textwidth}
      \begin{subfigure}{.33\textwidth}%
        \vspace{0.5cm}
        \includegraphics[width=\linewidth]{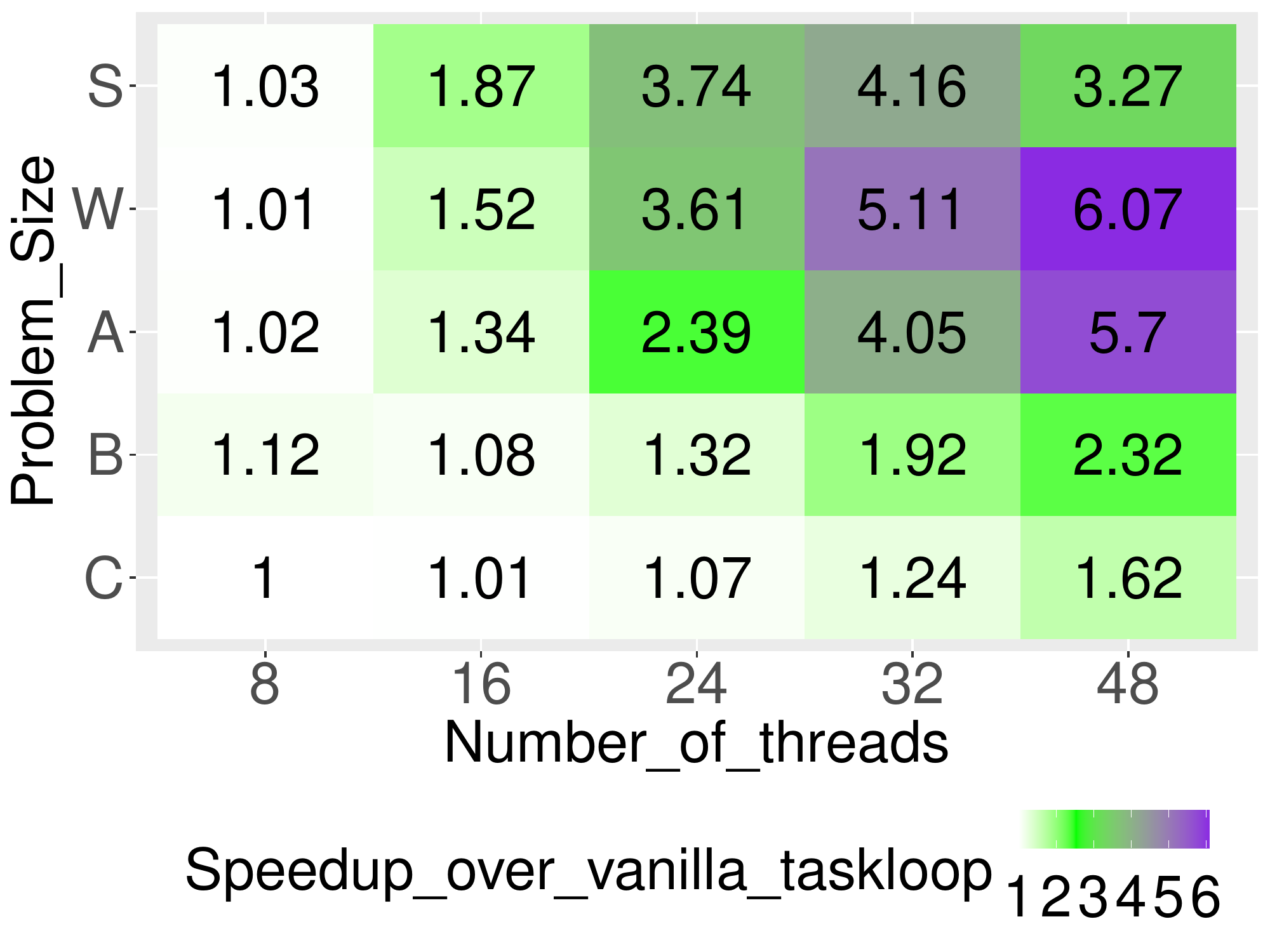}
        \vspace{-0.7cm}
        \caption{NAS\_CG}
        \label{fig:NAS_CG_static_heatmap}
      \end{subfigure}
      \hfill
      \begin{subfigure}{.33\textwidth}%
        \vspace{0.5cm}
        \includegraphics[width=\linewidth]{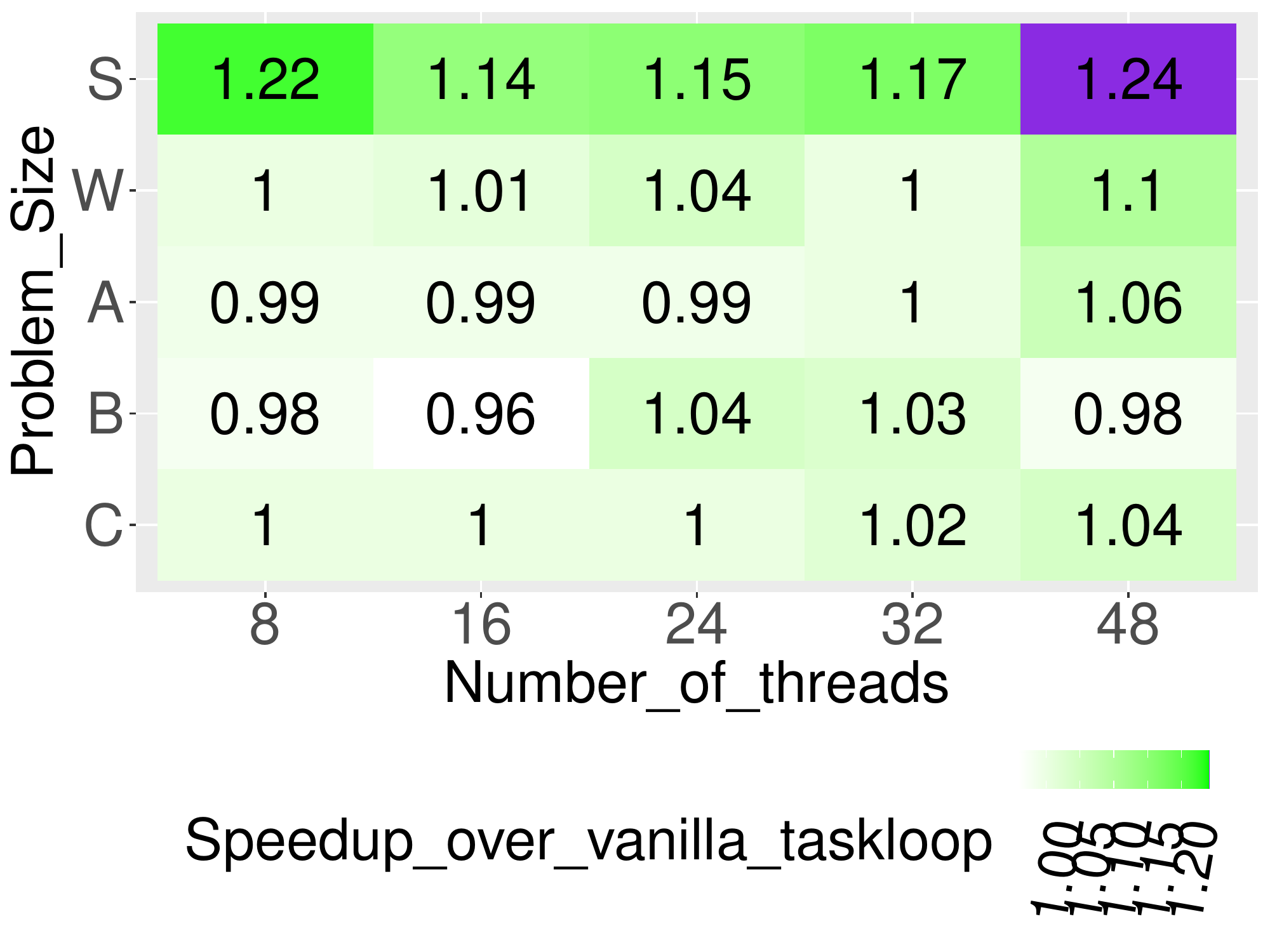}
        \vspace{-0.7cm}
        \caption{NAS\_FT}
        \label{fig:NAS_FT_static_heatmap}
      \end{subfigure}
      \hfill
      \begin{subfigure}{.33\textwidth}%
        \vspace{0.5cm}
        \includegraphics[width=\linewidth]{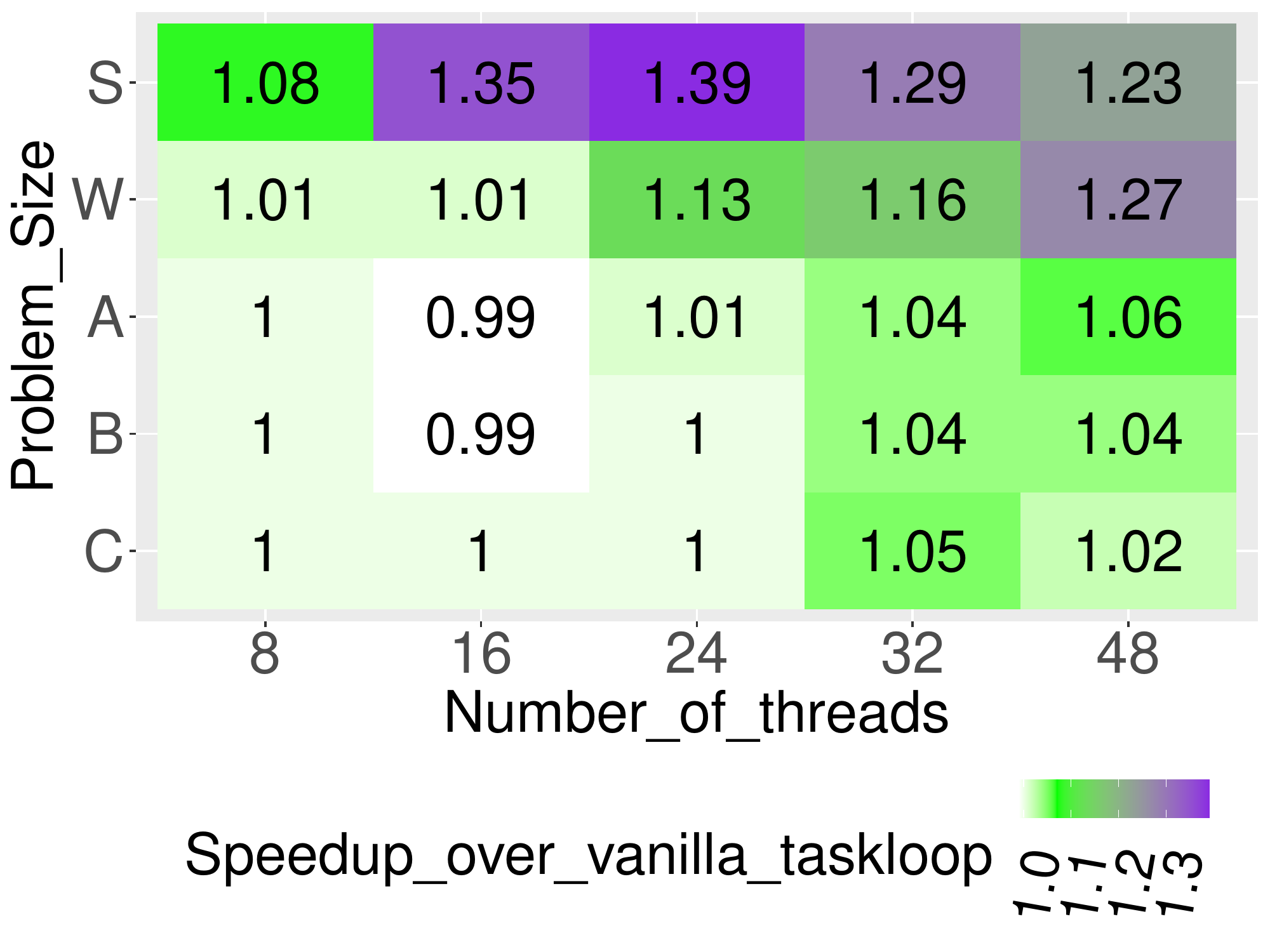}
        \vspace{-0.7cm}
        \caption{NAS\_BT}
        \label{fig:NAS_BT_static_heatmap}
      \end{subfigure}
      \hfill
      \begin{subfigure}{.33\textwidth}%
        \vspace{0.3cm}
        \includegraphics[width=\linewidth]{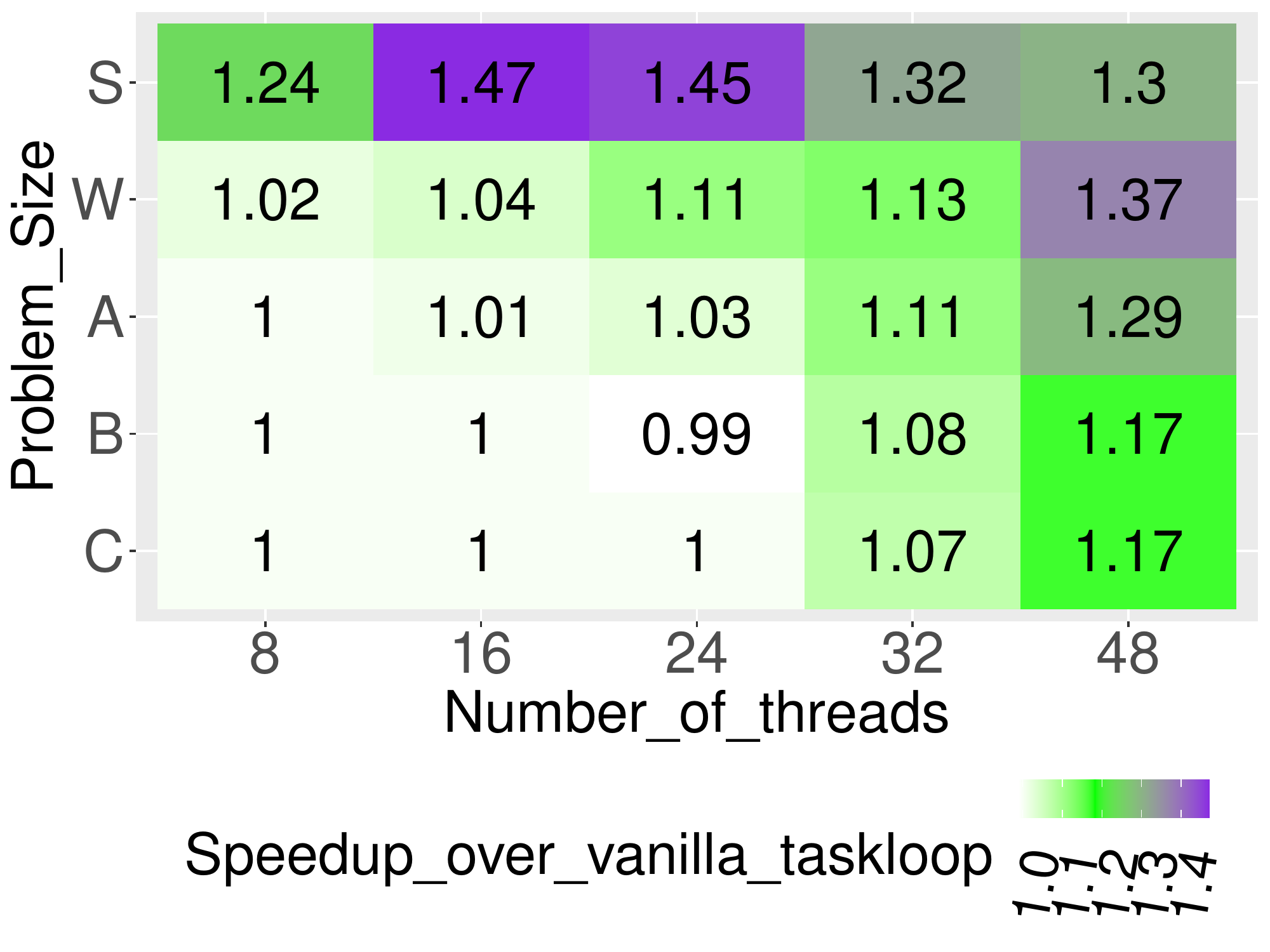}
        \vspace{-0.7cm}
        \caption{NAS\_SP}
        \label{fig:NAS_SP_static_heatmap}
      \end{subfigure}
      \hfill
      \begin{subfigure}{.33\textwidth}%
        \vspace{0.3cm}
        \includegraphics[width=\linewidth]{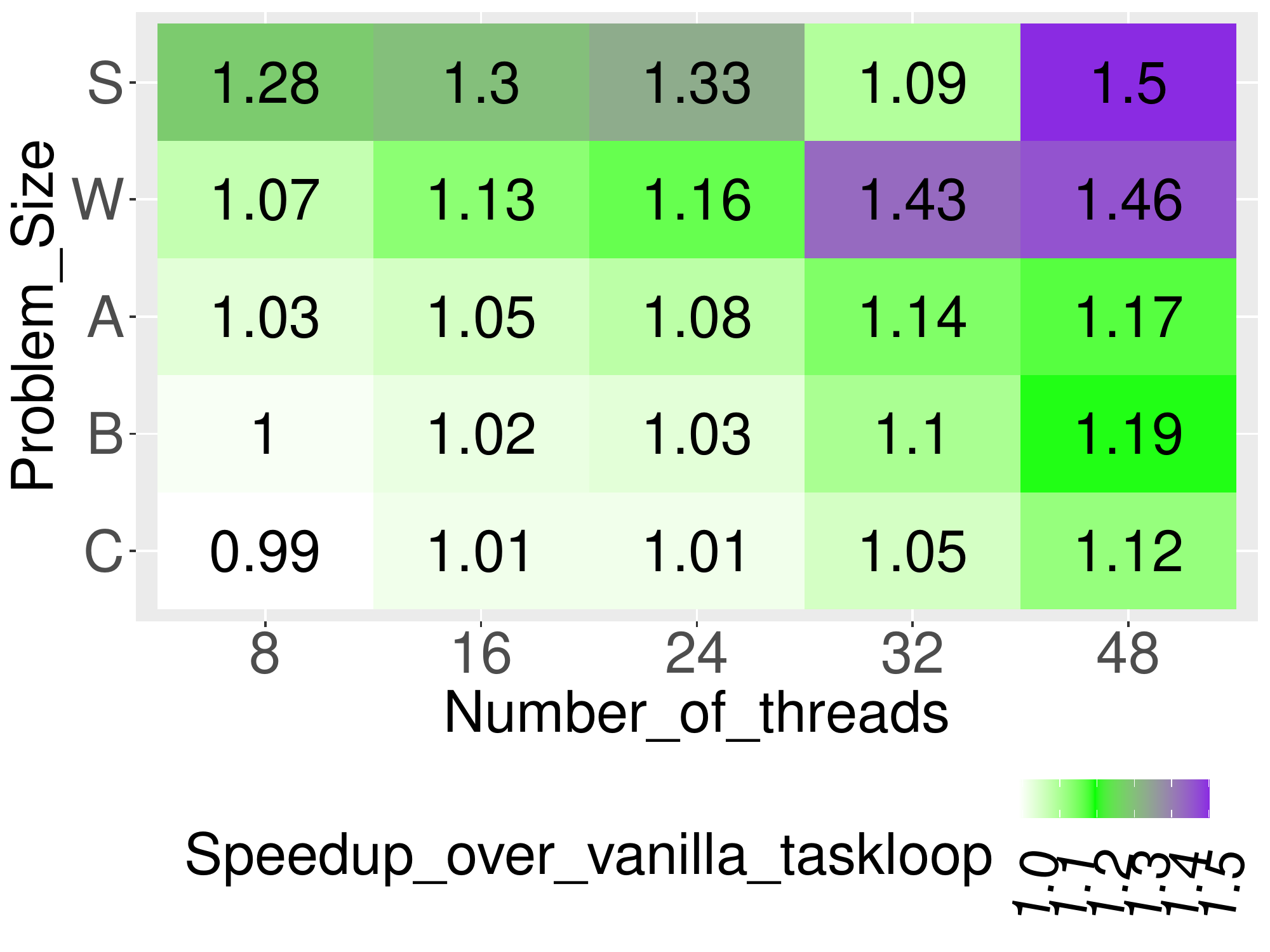}
        \vspace{-0.7cm}
        \caption{NAS\_LU}
        \label{fig:NAS_LU_static_heatmap}
      \end{subfigure}
      \hfill
      \begin{subfigure}{.33\textwidth}%
        \vspace{0.3cm}
        \includegraphics[width=\linewidth]{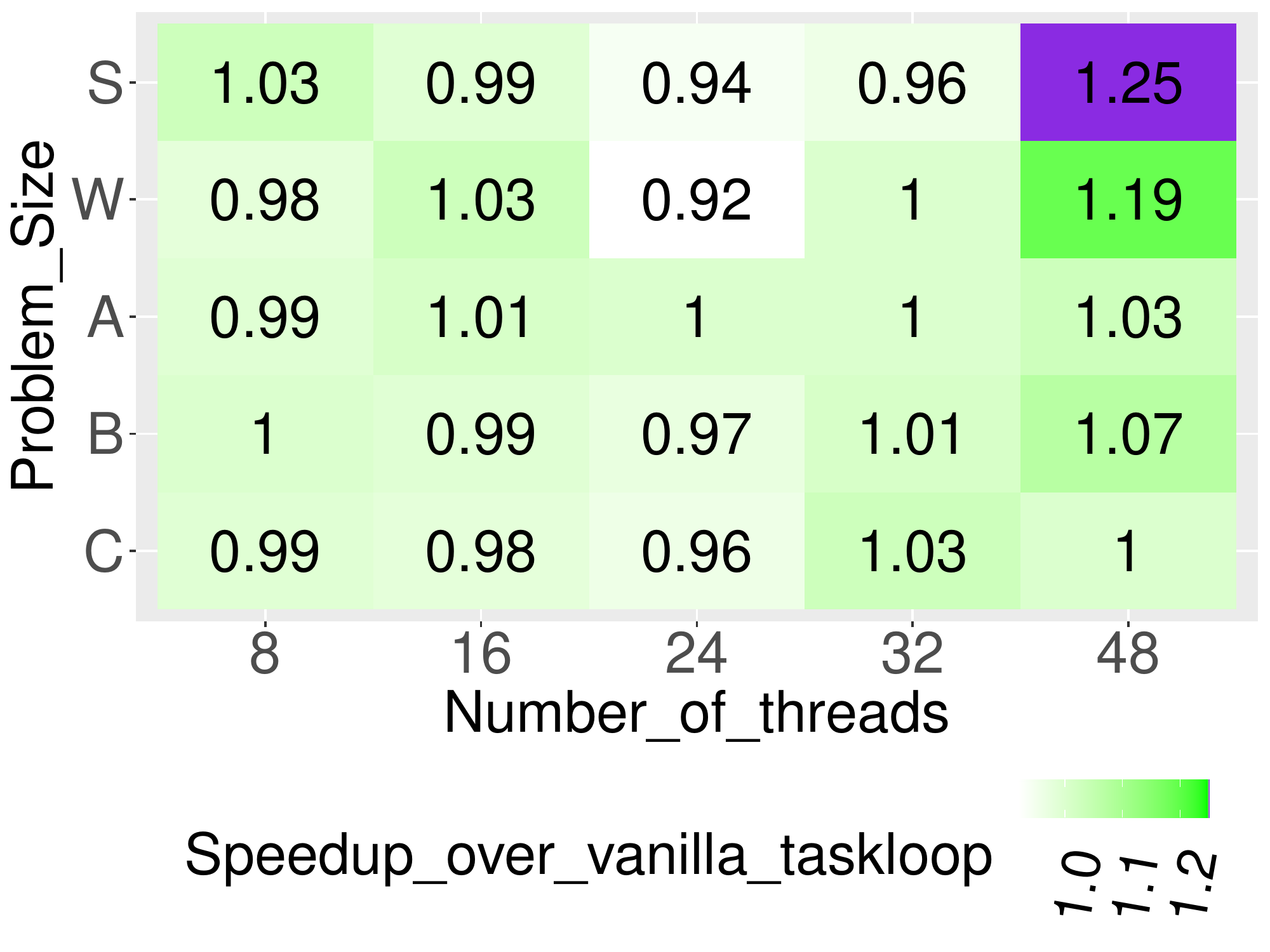}
        \vspace{-0.7cm}
        \caption{NAS\_EP}
        \label{fig:NAS_EP_static_heatmap}
      \end{subfigure}
      \vspace{-0.2cm}
      \caption{Normalized speedup of taskgraph compared to original \textit{taskloop} construct.}
      \label{fig:NAS_heatmaps}
      \end{minipage}
\end{figure*}

To verify that the reduction in the overhead reported above remains valid in 
more complex applications, we ran all benchmarks listed in 
Section~\ref{sec:experimentalSetup} with different settings. The results are 
reported in Figure~\ref{fig:unstruct_heatmaps} for unstructured parallelism 
(using the \texttt{task [depend]} construct), and Figure~\ref{fig:NAS_heatmaps} 
for structured parallelism (using the \texttt{taskloop} construct).

For all heatmaps in these two figures, the y-axis represents task granularity 
(large tasks at the bottom and small tasks on top). As previously stated, all 
benchmarks (except for the Nbody simulation) in 
Figure~\ref{fig:unstruct_heatmaps} are divided in blocks, and they keep their 
problem size unchanged, while setting different number of blocks to modify task 
granularity. For the Nbody simulation, the task workload is determined by 
setting different number of interacting particles. The numbers in the cells 
represent the speedup of Taskgraph execution over original \textit{task} (or 
\textit{taskloop}) execution computed as $\frac{Time\_task(or\ 
taskloop)}{Time\_Taskgraph}$. The higher the values, the faster Taskgraph 
execution is compared to \textit{task} or \textit{taskloop}.

Different scenarios are tested for both unstructured and structured parallelism, 
including those where tasking overhead is not of concern, i.e., little number of 
threads (in this case, 8 on a machine of 48 physical cores) and large task 
granularity. We observe a negative impact of the Taskgraph execution in some 
cases, but they are mostly mild (from 1\% to 8\%). However, in cases where the 
task granularity is small and the thread contention is important (more than 32 
threads), Taskgraph shows significant improvements. As an example, when 
executing with 48 threads, the NAS\_CG benchmark with Taskgraph shows a speedup 
of at least 1.6x than the original \textit{taskloop}, to a maximum of more than 
6x faster after setting the problem size to W.

\subsection{What If We Record TDGs}
\label{sec:dorr}
In Section~\ref{sec:over_reduc} and Section~\ref{sec:peak}, we assume the TDG is 
already built at execution, either defined statically or recorded dynamically. 
For instance, we record the TDG at the warm-up iteration in the NAS bencharmark 
suite. 
In this section, instead, we analyze the performance of the runtime while 
taking into account the cost of recording a TDG. This case considers that the 
TDG cannot be built at compile-time due to lack of information (e.g. problem 
size, pointer addresses), which is the most common scenario in HPC 
applications. 

\begin{figure}[h!]
  \begin{minipage}{\columnwidth}
    \centering
\includegraphics[width=.95\columnwidth]{
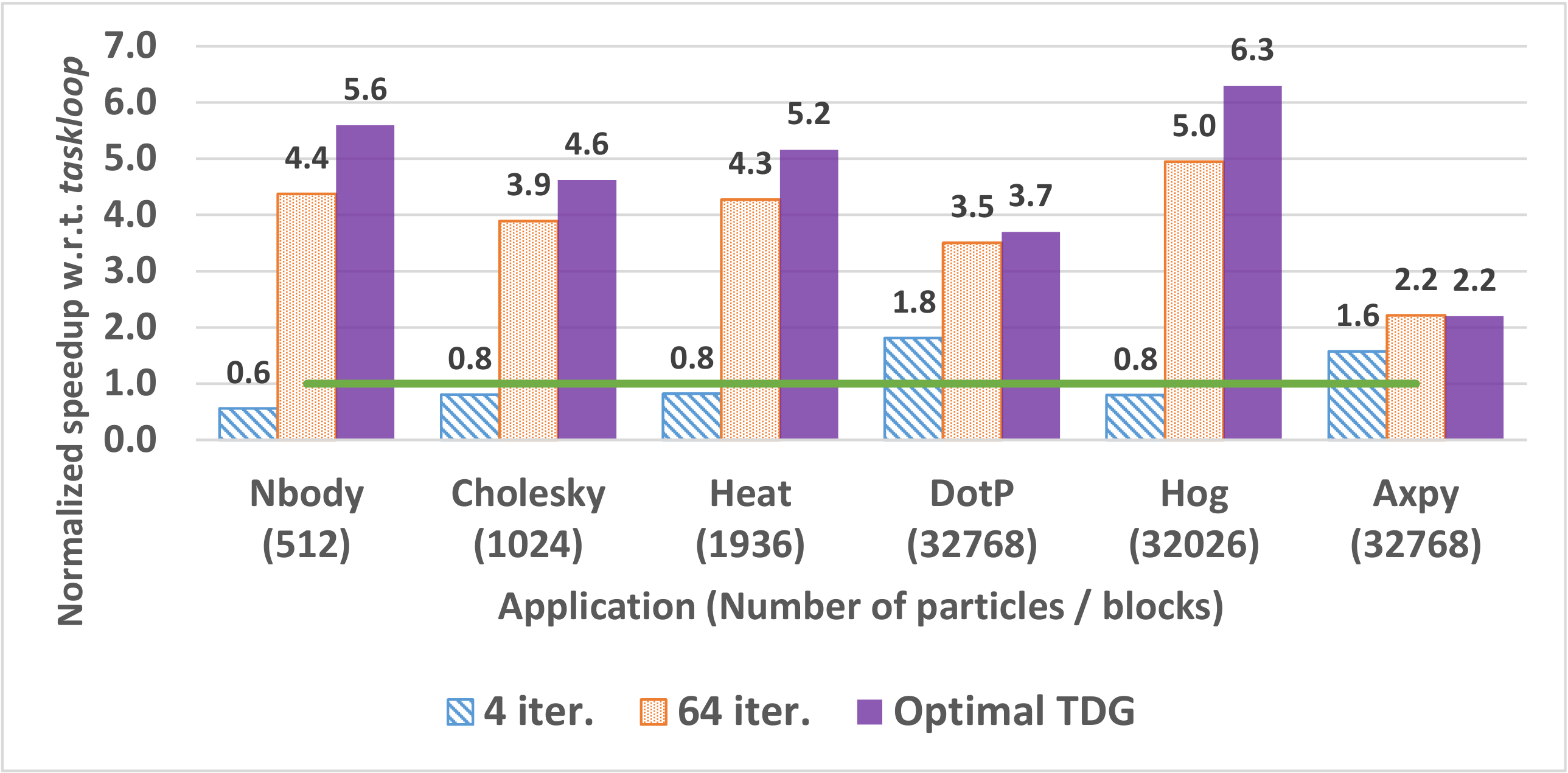}
    \subcaption{Smallest task granularity tested in 
Figure~\ref{fig:unstruct_heatmaps}.}
\includegraphics[width=.95\columnwidth]{
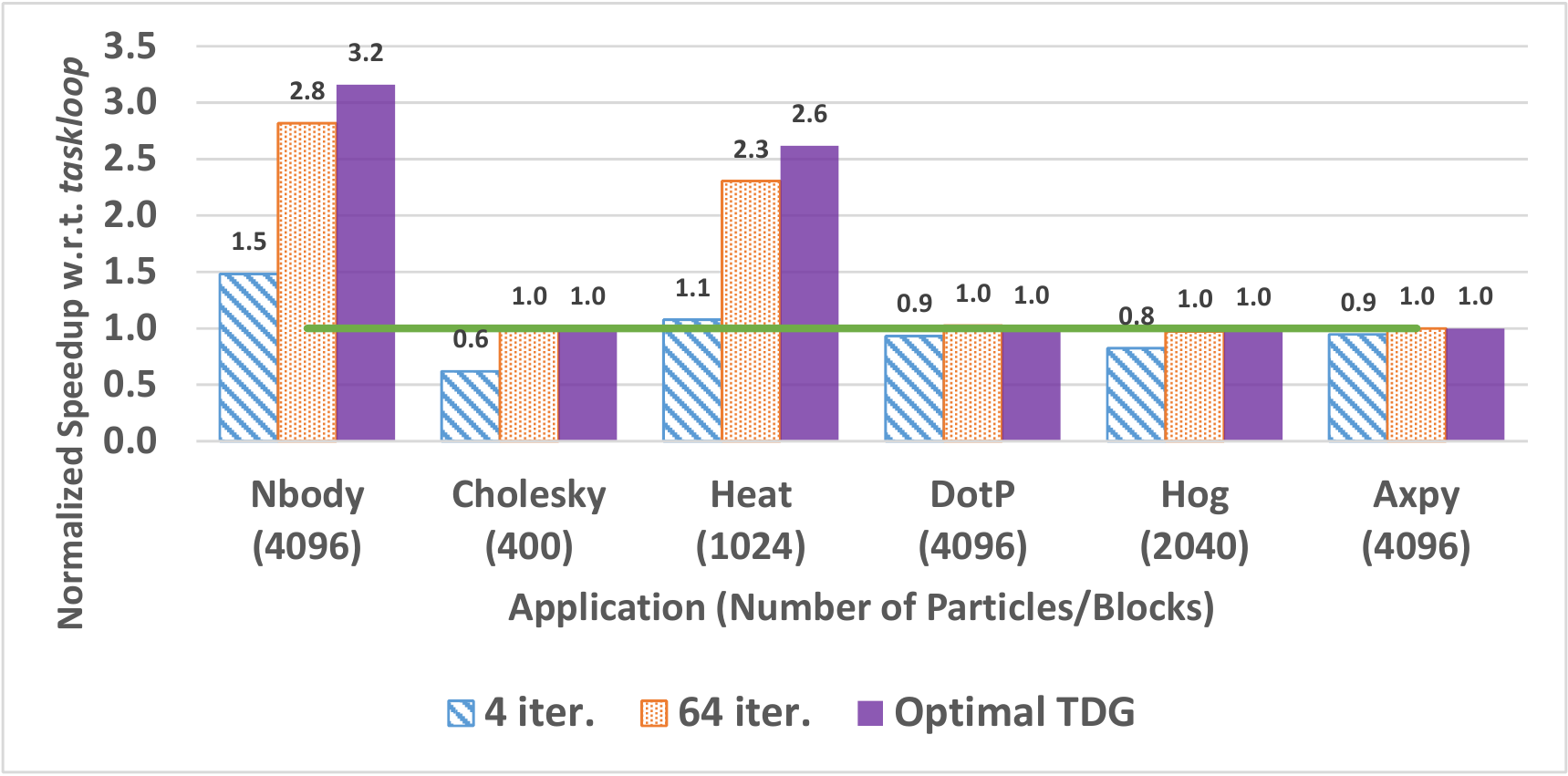}
    \subcaption{Largest task granularity tested in 
Figure~\ref{fig:unstruct_heatmaps}.}
  \end{minipage}
  \caption{Speedup of Taskgraph over original tasks. We set OMP\_NUM\_THREADS to 
48 (higher is better).}
  \label{fig:unstruct_bars}
\end{figure}
\begin{figure}[h!]
  \centering
  \begin{minipage}{\columnwidth}
  \centering  
\includegraphics[width=.95\columnwidth]{
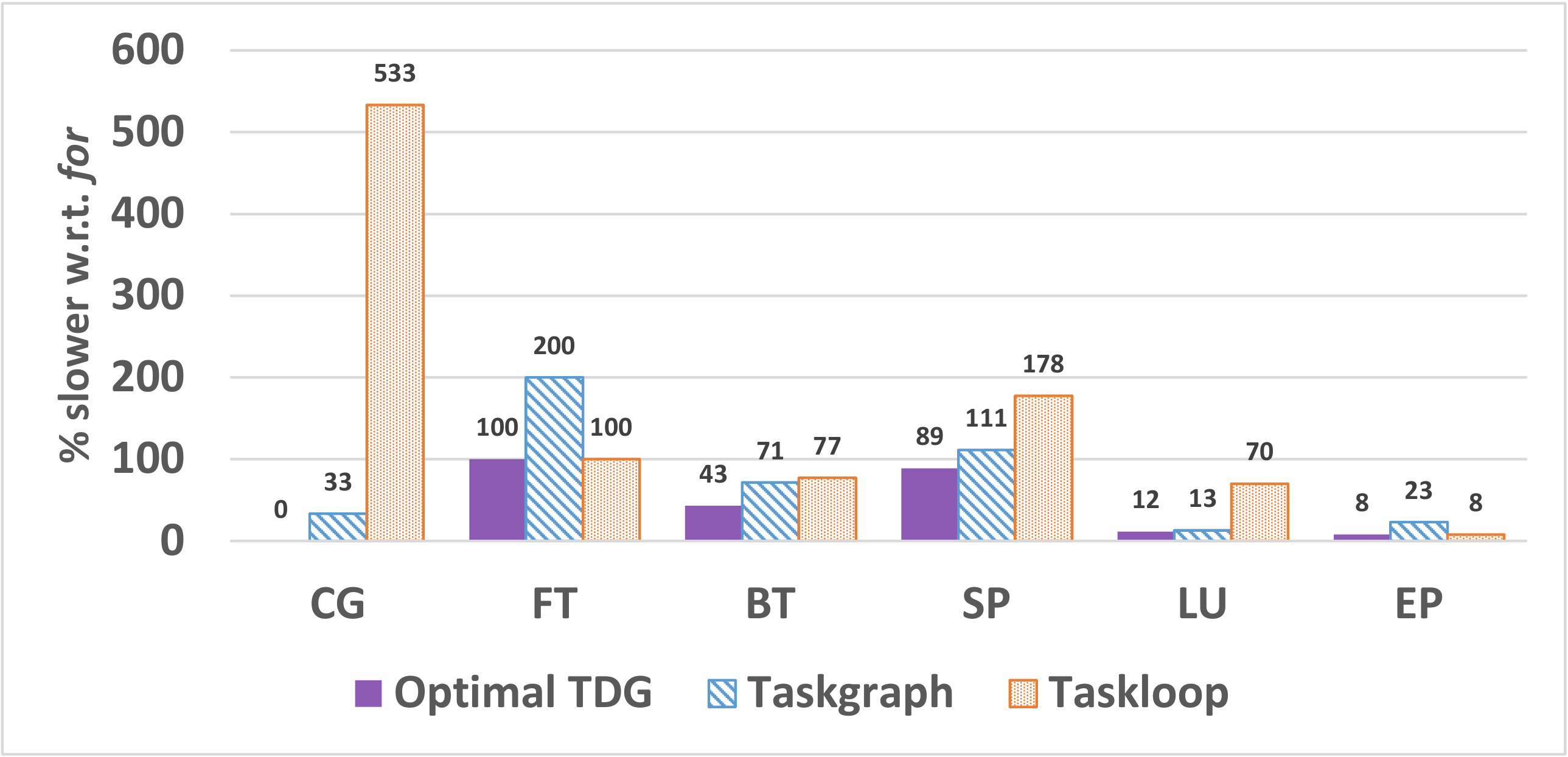}
    \subcaption{Problem size W.}
\includegraphics[width=.95\columnwidth]{
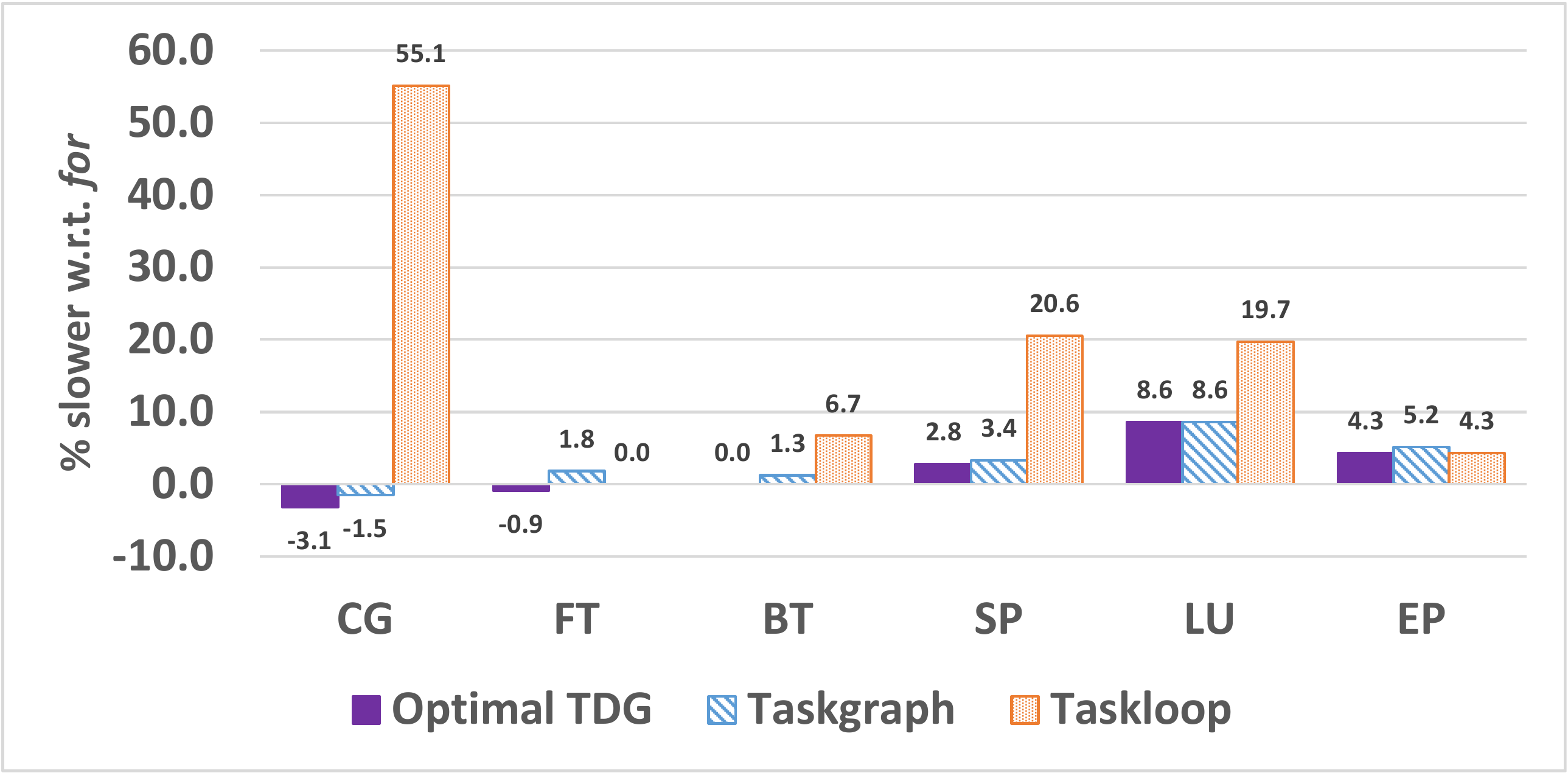}
    \subcaption{Problem size C.}
  \end{minipage}
  \caption{Overhead incurred by Taskgraph and \texttt{taskloop} after replacing 
\texttt{for} in the original NAS implementation, using 48 threads (lower is 
better).}
  \label{fig:NAS_bars}
\end{figure}

We set the number of threads to 48 and execute all benchmarks various times. 
The results corresponding to 4 and 64 iterations are reported in 
Figures~\ref{fig:unstruct_bars} and~\ref{fig:NAS_bars}.
The former shows the speedup of Taskgraph over the original tasking model. 
Values below 1 mean the Taskgraph execution is slower than the vanilla 
implementation. This is the case for some applications when the number of 
iterations is low, and the recording overhead cannot be amortized. This overhead 
is gradually mitigated when we increase the number of iterations. After 64 
iterations, the speedup of the Taskgraph framework is close to the optimal one 
presented in Figure~\ref{fig:unstruct_heatmaps}.

To evaluate the framework with structured tasks, we compare the performance of 
\texttt{taskloop} and Taskgraph against the original implementation of the NAS 
benchmarks using \texttt{for} constructs. The results reported in 
Figure~\ref{fig:NAS_bars} are computed as 
$\frac{(Measured\_Time-Time\_for)}{Time\_for}$, where $Measured\_Time$ includes 
the overhead of recording (lower values are better).

One can notice that for CG and FT with problem size W, Taskgraph is notably
slower than the optimal TDG, because their number of iterations are particularly 
low i.e., 15 and 6 respectively, and do not allow to amortize the recording 
overhead. Once we increase the problem size, the number of iterations start to 
grow and the differences between Taskgraph and Optimal TDG become minuscule. In 
the case of larger problem size, we can also notice that Taskgraph has a 
performance closer to the thread model if compared to \textit{taskloop}. For CG, 
\texttt{taskloop} is about 55\% slower than \texttt{for}, however, Taskgraph 
can potentially be 3\% faster than the thread model.

\subsection{Proof of concept: GCC implementation}

\begin{figure}[h!]
  \centering
    \begin{minipage}[t]{\columnwidth}
      \centering
\includegraphics[width=.95\columnwidth]{
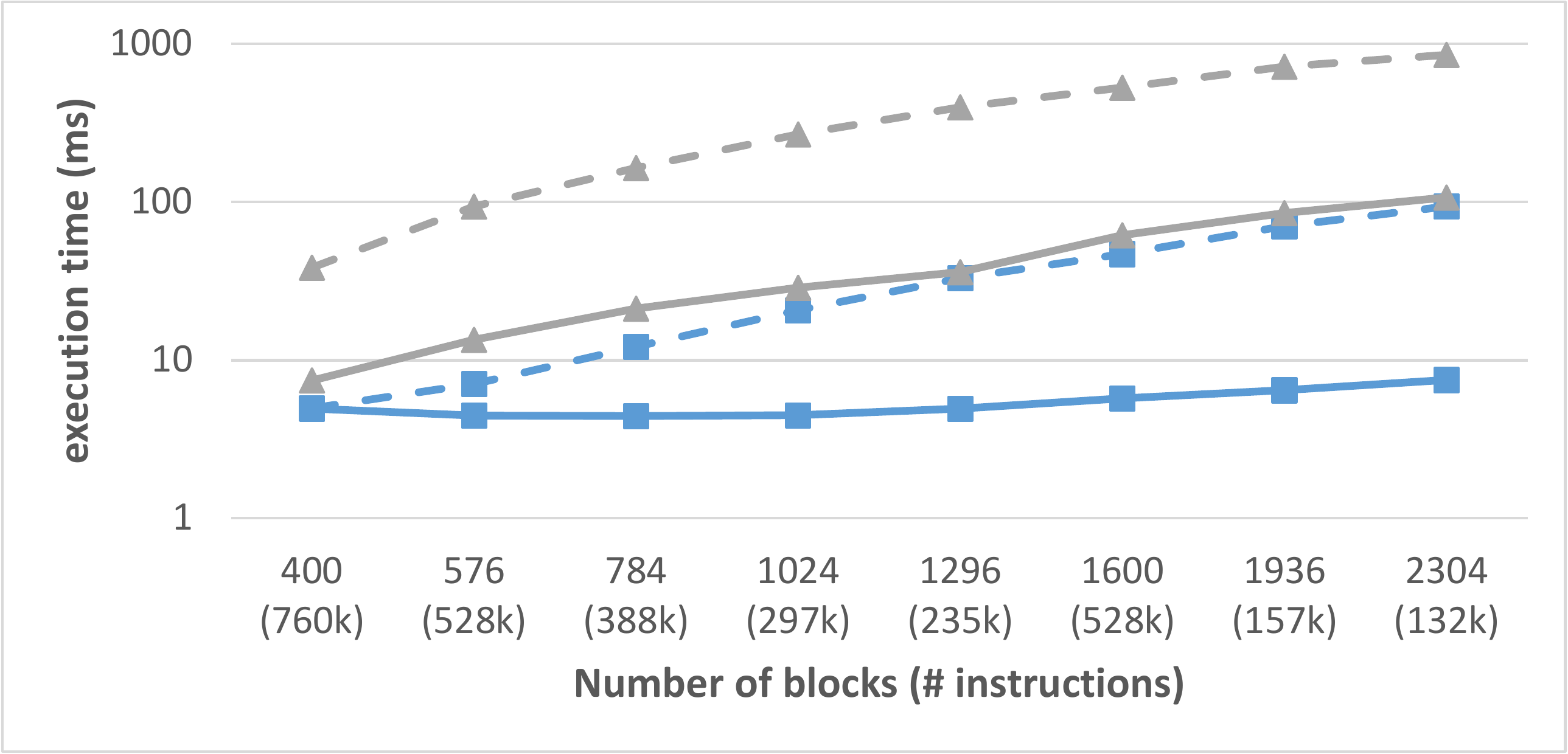}
      \subcaption{Cholesky matrix decomposition.}
      \label{fig:normalized_llvm}
    \end{minipage}
    \begin{minipage}[t]{.95\columnwidth}
      \centering
      
\includegraphics[width=1\columnwidth]{
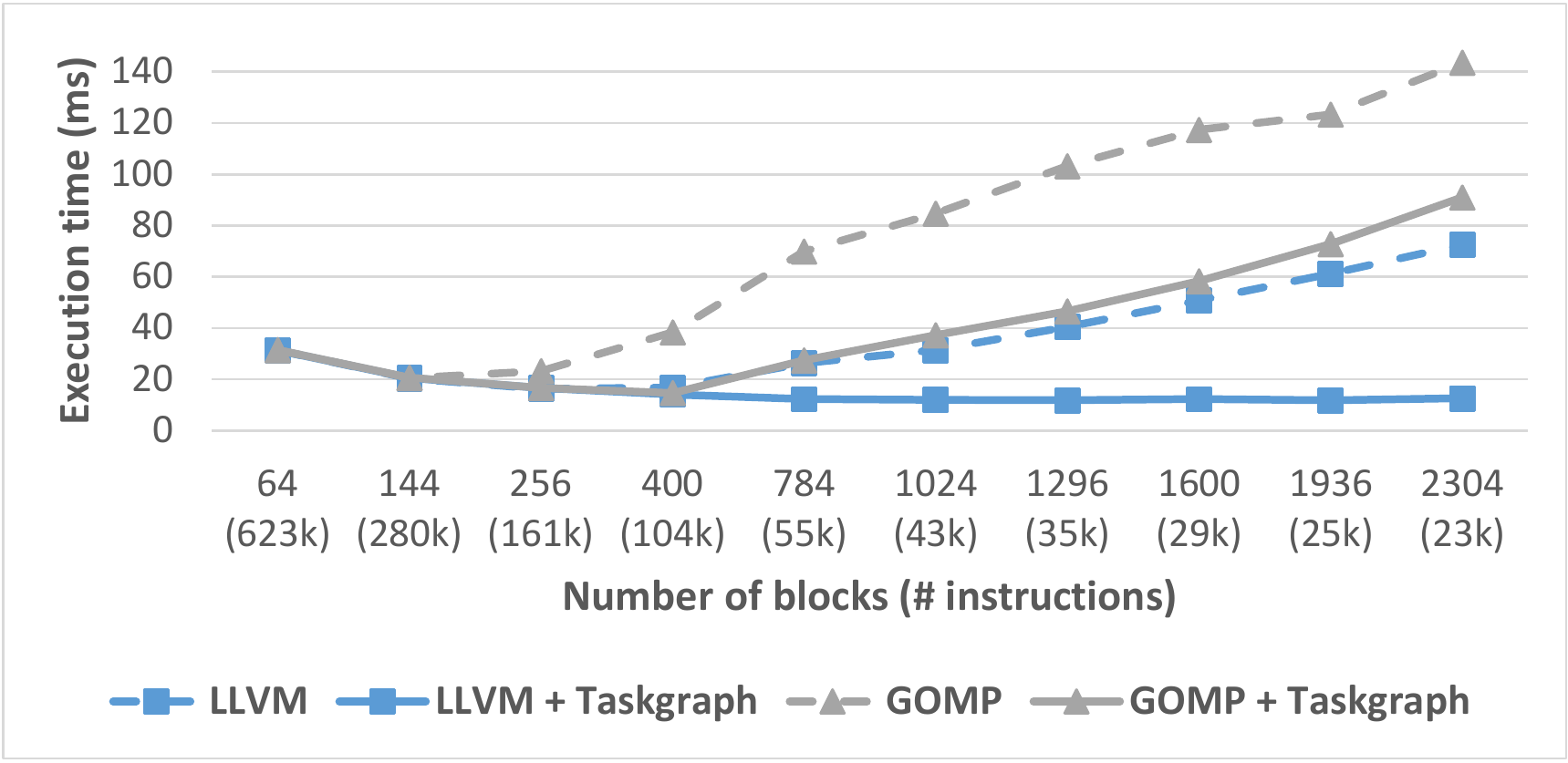}
      \subcaption{Heat diffusion simulator.}
      \label{fig:normalized_gcc}
    \end{minipage}
  \caption{A breakdown of the original and the taskgraph runtime systems 
performance using different task granularities and 48 threads.}
  \label{fig:breakdown}
\end{figure}

To prove the portability of the results obtained with LLVM, we have implemented 
support for Taskgraph execution in GCC's OpenMP library. 
Figure~\ref{fig:breakdown} shows that, for both Cholesky decomposition and Heat 
propagation algorithms, and all tested task granularities, runtime libraries 
that adopted \texttt{Taskgraph} deliver either similar performance as the 
original libraries (when tasks are of coarse grain) or shorter execution time 
(for fine-grained tasks). More interestingly, the Taskgraph framework shows 
reasonable stability in the execution time when the granularity of the tasks is 
drastically reduced (e.g., up to 23K in the Heat benchmark). The same trend has 
been observed for all benchmarks.

\section{Conclusions and future work}
\label{sec:ccl}
This paper presents a new OpenMP task-based framework, named \texttt{Taskgraph}, 
for optimised parallel execution of the OpenMP tasking model, with reduced 
runtime overheads related to the OpenMP task management, i.e., contention and 
parallel orchestration. Specifically, the framework addresses the optimization 
of fully-taskified regions of code, that can be represented as a Task 
Dependency Graph (TDG). To that end, it covers regions that can be fully 
expanded at compile-time, and also regions which TDG's shape is decided at  
run-time. To exploit the latter, a record-and-replay mechanism is implemented.
Typically, applications from the embedded computing domain can exploit static 
TDGs, while HPC systems running kernels repeatedly can exploit dynamic TDGs.

Our evaluation shows the benefits of the optimizations in the execution enabled 
by both static and dynamic TDGs, allowing to exploit applications with higher 
degree of finer-grained tasks. Specifically, compared to the native 
\textit{task} and \textit{taskloop} environments in LLVM, we achieved a 
speedup of up to 6x when applied Taskgraph framework.
Also the scalability of the taskgraph framework is better, showing 
enhanced performance for a wider range of configurations.

One limitation of Taskgraph framework is that it requires a fully taskified 
region, i.e., in the opposite case, users must manually taskify sequential code 
with correct dependencies so that the TDG includes the workload of the entire 
region. To tackle this limitation automatically, we consider implementing a 
compiler analysis capable of identifying a smaller region that fulfills this 
condition, and apply the framework to a subset of tasks, instead of the entire 
\texttt{taskgraph} region. We then link this optimized task subset to the rest 
of the region by input and output dependencies.

Although the OpenMP specification has been moving from prescriptive techniques 
based on the thread model into descriptive techniques based on the tasking 
model, it is hard to find real-world applications exploiting the latter model. 
There are some test suites dedicated to the OpenMP preliminary tasking 
model~\cite{duran2009barcelona} and OpenMP tasks with 
dependencies~\cite{virouleau2014evaluation}, but they do not show some 
characteristics assumed in this work. As a result, the current framework is  
tested in specific kernels from both the EC and the HPC domains. Full 
applications are already targeted for further testing the framework, including 
examples coming from the EC domain (e.g., WATERS challenges 
\cite{waters2018challenge,waters2019challenge}) and from the HPC domain (e.g., 
sLAS linear algebra solver \cite{valero2020slass} and the Quantum ESPRESSO 
material modeling tool \cite{giannozzi2009quantum}).


%



\ifCLASSOPTIONcompsoc
  \section*{Acknowledgments}
\else
  \section*{Acknowledgment}
\fi
This work has been supported by the Generalitat de Catalunya project RESPECT under record
no. 2021 PROD 00179, and the EU H2020 project AMPERE under the grant agreement no. 871669.
\ifCLASSOPTIONcaptionsoff
  \newpage
\fi



\bibliographystyle{IEEEtran}
\bibliography{IEEEabrv,./paper}
%



%

\begin{IEEEbiography}[{\includegraphics[width=1in,height=1.25in,clip,keepaspectratio]{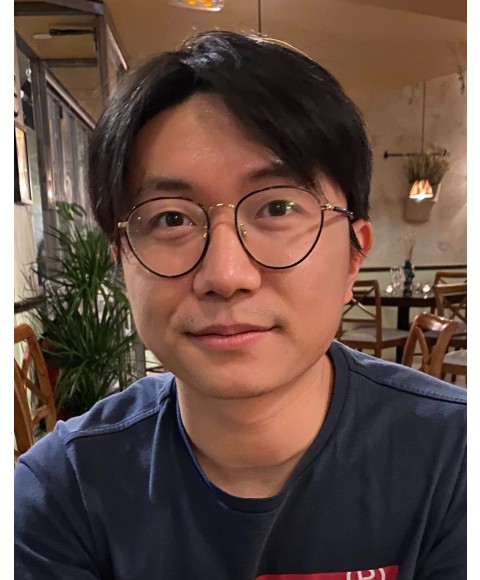}}]{Chenle Yu}
received his MS degree in High Performance Computing (HPC) and 
Cryptography in 2018 at Sorbonne University, France. He is now pursuing his 
Ph.D. degree in the Computer Architecture department at Polytechnic 
University of Catalonia (UPC), Spain. His research is now focused on parallel
programming models, more specifically, on-node parallelism using OpenMP 
and its compiler, runtime system, together with the extension of current
OpenMP offloading feature.
\end{IEEEbiography}

\begin{IEEEbiography}[{\includegraphics[width=1in,height=1.25in,clip,keepaspectratio]{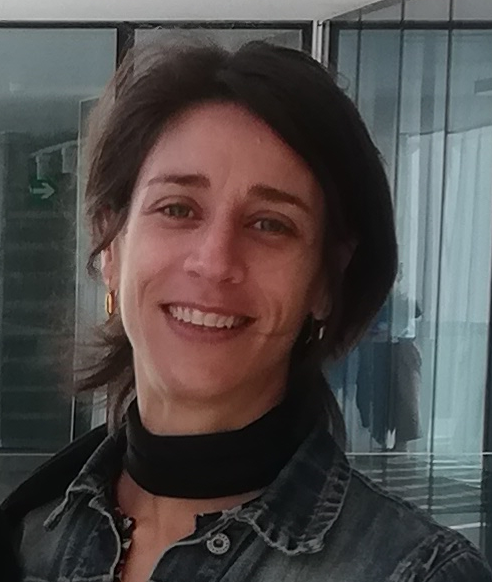}}]{Sara Royuela}
Dr. Sara Royuela is a senior researcher in the Predictable Parallel Computing
group at BSC leading the activities related to parallel productivity and
dependability in multi-core and acceleration devices. Sara coordinates the
RESPECT project, is WP leader in AMPERE and participates in Rising STARS, all
related to the evolution of HPC parallel programming models towards performance,
reliability, predictability and interoperability. Sara is an expert on
compilation techniques for parallel systems.
\end{IEEEbiography}


\begin{IEEEbiography}[{\includegraphics[width=1in,height=1.25in,clip,keepaspectratio]{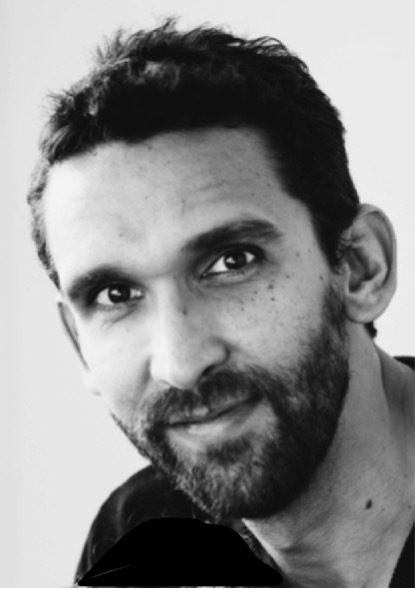}}]{Eduardo Quiñones}
Dr. Eduardo Quiñones is a senior researcher, head of the Predictable Parallel
Computing group at BSC and member of HiPEAC. He has coordinated CLASS and
ELASTIC European projects, and coordinates AMPERE, all related to the use of
HPC parallel programming models across the computing continuum. He coordinates
BSC efforts in DeepHealth and the RISE RisingSTARS. His research focuses on
compiler techniques, many-core architectures and distributed computing for
safety-critical systems.
\end{IEEEbiography}




\end{document}